\documentclass[preprint,12pt]{emulateapj-rtx4}
\usepackage[varg]{txfonts}
\usepackage{graphicx}
\usepackage{booktabs}
\usepackage{afterpage}
\usepackage{changepage}
\usepackage{threeparttable}
\usepackage{chngcntr}
\bibpunct{(}{)}{;}{a}{}{,} 
\usepackage{color}
\usepackage[caption=false]{subfig}
\usepackage[colorlinks=true,citecolor=blue,urlcolor=green]{hyperref}

\usepackage{longtable}

\newlength{\offsetpage}
{\end{adjustwidth}}

\begin{document}

\title{The orbital eccentricity of small planet systems}

\hyphenation{Kepler}

\author{Vincent~Van~Eylen$^{1,2,3}$, Simon~Albrecht$^3$, Xu Huang$^{4}$, Mariah G.\ MacDonald$^{5}$, Rebekah I.\ Dawson$^{5}$, Maxwell X.\ Cai$^{2}$,\\ Daniel Foreman-Mackey$^{6}$, Mia S.\ Lundkvist$^{3,7}$, Victor Silva Aguirre$^3$, Ignas Snellen$^{2}$, Joshua N.\ Winn$^1$}

\affil{	
$^1$ Department of Astrophysical Sciences, Princeton University, 4 Ivy Lane, Princeton, NJ 08540, USA\\
$^2$ Leiden Observatory, Leiden University, 2333CA Leiden, The Netherlands\\
$^3$ Stellar Astrophysics Centre, Department of Physics and Astronomy, Aarhus University, Ny Munkegade 120, \\
DK-8000 Aarhus C, Denmark \\
$^4$ MIT Kavli Institute for Astrophysics and Space Research, 70 Vassar St., Cambridge, MA 02139\\
$^5$ Department of Astronomy \& Astrophysics, and Center for Exoplanets and Habitable Worlds,\\
525 Davey Lab, The Pennsylvania State University, University Park, PA, 16802, USA \\
$^6$ Center for Computational Astrophysics, Flatiron Institute, 162 Fifth Avenue, New York, NY 10010, USA\\
$^7$ Zentrum f\"ur Astronomie der Universit\"at Heidelberg, Landessternwarte, K\"onigstuhl 12, 69117 Heidelberg, Germany\\
}

\email{vaneylen@astro.princeton.edu}

\shorttitle{Eccentricity distribution of transiting planets}
\shortauthors{Van Eylen et al.}

\received{receipt date}
\revised{revision date}

\begin{abstract}{
We determine the orbital eccentricities of individual small
\textit{Kepler} planets, through a combination of asteroseismology and
transit light-curve analysis.  We are able to constrain the
eccentricities of 51 systems with a single transiting planet, which
supplement our previous measurements of 66 planets in multi-planet
systems.
Through a Bayesian hierarchical analysis, we find evidence that
systems with only one detected transiting planet have a different
eccentricity distribution than systems with multiple detected
transiting planets.  The eccentricity distribution of the
single-transiting systems is well described by the positive half of a
zero-mean Gaussian distribution with a dispersion $\sigma_e = 0.32 \pm 0.06$, while the multiple-transit systems are consistent with $\sigma_e = 0.083^{+0.015}_{-0.020}$. A mixture model suggests a fraction of $0.76^{+0.21}_{-0.12}$ of single-transiting systems have a moderate eccentricity, represented by a Rayleigh distribution that peaks at $0.26^{+0.04}_{-0.06}$.
This finding may reflect differences in the formation pathways of
systems with different numbers of transiting planets.  We investigate the possibility
that eccentricities are ``self-excited'' in closely packed
planetary systems, as well as the influence of long-period giant companion planets. We find that both mechanisms can qualitatively explain the observations. 
We do not find any evidence for a
correlation between eccentricity and stellar metallicity, as has been
seen for giant planets.  Neither do we find any evidence that orbital
eccentricity is linked to the detection of a companion star.
Along with this paper we make available all of the parameters and uncertainties
in the eccentricity distributions, as well as the properties of
individual systems, for use in future studies.
}
\end{abstract}

\maketitle
 
\section{Introduction}

The known planets in the solar system have nearly circular orbits, with a mean
eccentricity ($e$) of 0.04.
Gas giant exoplanets show a wide range
of eccentricities \citep[e.g.][]{butler2006}.  The current record holder
for the highest eccentricity is HD~20782b, with $e=0.956\pm 0.004$ \citep{kane2016}.
Smaller exoplanets are not as well explored. It would be interesting
to constrain their eccentricity distribution, in order to gain
clues about their formation and evolution. A number of physical processes can damp or
excite orbital eccentricities
\citep[e.g.][]{rasio1996,fabrycky2007,chatterjee2008,juric2008,ford2008}.

However, measuring eccentricities for small planets can be difficult.
The radial velocity (RV) signal associated with a small planet is
small, and in many cases undetectable with current instruments.  Even
if the RV signal can be detected, the eccentricity is one of the most
difficult parameters to constrain, and is often assumed to be zero
unless the data are of unusually high quality
\citep[e.g.][]{marcy2014}. Small planets can be detected with the
transit method, but the mere detection of transits usually does not
provide enough information to determine the orbital eccentricity.  For
short-period planets, the relative timing of transits and occultations
can be used to constrain the eccentricity \citep{shabram2016}.  When
transit timing variations are detectable, they too can sometimes be
used to infer the underlying eccentricity \citep[see, e.g.,][]{hadden2014}.

A method with wider applicability relies on accurate determinations of
the transit duration, the transit impact parameter, and the stellar
mean density.  Many variations on this technique have been described
in the literature \citep[see,
  e.g.,][]{ford2008,tingley2011,dawson2012,kipping2014,vaneylen2015,xie2016}.
Many previous attempts to perform this type of analysis on {\it
  Kepler} planets have been frustrated by the lack of accurate and
unbiased estimates of the stellar mean density \citep[see,
  e.g.,][]{sliski2014,plavchan2014,rowe2014}.

Using a subsample of \textit{Kepler} systems with stellar mean
densities derived from asteroseismology, \cite{vaneylen2015} derived
the eccentricity of 74 planets in multi-planet systems, and found that
the data are compatible with a Rayleigh distribution peaking at
$\sigma_e = 0.049 \pm 0.013$. \cite{xie2016} studied the eccentricity
distribution of a larger sample of \textit{Kepler} planets. They used
homogeneously derived spectroscopic stellar densities from the Large Sky Area Multi-Object Fibre Spectroscopic Telescope 
(LAMOST) survey, which are less precise than asteroseismic
stellar densities, typically by up to an order of magnitude, and were the dominant source of uncertainty. They found that systems with a single detected
transiting planet have an average eccentricity of $\approx$\,0.3, and
that systems with multiple detected transiting planets have a
significantly lower mean eccentricity of $0.04^{+0.03}_{-0.04}$.

Here, we use asteroseismically derived stellar mean densities to
derive eccentricities for individual transiting planets in {\it
  Kepler} systems with only a single detected transiting planet.  We
combine the results with our previous results for multi-planet systems
\citep{vaneylen2015}, to investigate any possible differences between
these two populations, as well as to search for any correlations
between eccentricity and other planetary and stellar parameters.
In Section~\ref{sec:methods}, we describe the planet sample and
analysis methods. The results are presented in
Section~\ref{sec:results}. In Section~\ref{sec:interpretation}, we
interpret the findings and compare them with planet formation and
evolution models. We discuss our findings and compare them to previous
work in Section~\ref{sec:discussion}, and draw conclusions in
Section~\ref{sec:conclusion}.

\section{Methods}
\label{sec:methods}

\subsection{Sample selection}
\label{sec:sampleselection}

To ensure accurate and precise stellar parameters, in particular mean stellar densities, we select a sample of planet candidates in single-planet systems for which the stellar parameters were determined in a homogeneous asteroseismic analysis \citep{lundkvist2016}. This sample consists of 64 candidates, of which 5 systems (KOI-1283, KOI-2312, KOI-3194, KOI-5578, and KOI-5665) were excluded because fewer than six months of \textit{Kepler} short-cadence data are available and we found that these systems had too few observed transits to determine precise transit parameters. KOI-3202 was removed because only a single transit was observed in short cadence. KOI-5782 has only two transits observed in short cadence and is also excluded. KOI-2659, KOI-4198, and KOI-5086 have been flagged as false positives\footnote{See the NASA Exoplanet Archive, \url{https://exoplanetarchive.ipac.caltech.edu}.} and removed from the sample. KOI-2720 shows a very strong spot signal, complicating the precise modeling of the transit, and we exclude this system from further analysis. The remaining sample consists of relatively bright stars (with a median \textit{Kepler} magnitude of $11.5$) with precisely determined stellar densities (with a median uncertainty of 4.5\%). 
The accuracy of stellar properties have been tested using independent constraints from interferometry \citep{silvaaguirre2017}, parallaxes \citep{silvaaguirre2012,huber2017,sahlholdt2018}, as well as common ages from binary systems \citep{silvaaguirre2017}. The level of accuracy obtained for stellar radii, as they are determined in \cite{lundkvist2016} and used here, is better than 5\% \citep{sahlholdt2018}. The latest Gaia release indicates even better agreement with parallax, down to a level of 3\% \citep{sahlholdt2018b}. It is more difficult to assess the accuracy for mass determinations, due to lack of independent measurements, but the agreement of asteroseismic ages of clusters with those determined by isochrone fitting suggest accuracies better than 10\% in mass \citep[e.g.][]{stello2016,miglio2016,brogaard2018}.

Out of the 53 stars in our sample, 36 have a validated or confirmed planet, while 17 contain detected `planet candidates'. A study of the average \textit{Kepler} false positive rate by \cite{morton2011} find it to be below 10\% for most systems, and a further study by \cite{fressin2013} measured it to be 9.4\%. Similarly, a \textit{Spitzer} follow-up study of \textit{Kepler} planets found that at least 90\% of the \textit{Kepler} signals are planetary~\citep{desert2015}. Furthermore, our sample consists of very bright stars, which are observed to have a false positive rate several times lower than fainter stars \citep{desert2015}. 

\cite{morton2016} calculate false positive probabilities (FPPs) for individual KOIs, where values above 1\% are typically deemed insufficient to validate an individual system. We list the values for the unconfirmed planets in our sample in Table~\ref{tab:fpp_morton}. KOI-1537 and KOI-3165 have a FPP of 1, implying that they are likely false positives. We exclude these systems from our sample. All other candidates have low FPPs, suggesting our remaining sample has very few, if 
any, false positives: the combined FPP of all retained systems is 0.54, suggesting that our sample contains no more than one or two false positives.

While our sample consists exclusively of systems with a single detected transiting planet, some or all of these systems may nevertheless have additional undetected planets. In fact, in several cases, there is evidence of additional planets: six of the candidates exhibit TTVs (see Table~\ref{tab:eccsingle_ttvs}) and several of the other candidates have long-term RV variations indicative of distant planets. In this work, when we investigate `single planet' systems, this will normally mean planets with a single detected \textit{transiting} planet, and we use the term `single tranets', originally coined by \cite{tremaine2012}, to make this point more explicit when appropriate. 

\begin{table}[htbp]
\begin{center}
\footnotesize
\caption{The candidate planets in our sample that have not been previously validated or confirmed, and their false positive probability (FPP).\label{tab:fpp_morton}}
\begin{tabular}{lc}
KOI 	& FPP$^{1}$ \\
 \hline \hline
KOI-75b & 0.017 $\pm$ 0.027\\
KOI-92b & 0.091 $\pm$ 0.011\\
KOI-268b & 0.015 $\pm$ 0.003\\
KOI-269b & 0.023 $\pm$ 0.013\\
KOI-280b & 0.017 $\pm$ 0.003\\
KOI-288b & 0\\
KOI-319b & 0.041 $\pm$ 0.024\\
KOI-367b & 0.030 $\pm$ 0.025\\
KOI-974b & 0.0042 $\pm$ 0.0058\\
KOI-1537b & 1\\
KOI-1962b & 0.036 $\pm$ 0.004\\
KOI-1964b & 0.13 $\pm$ 0.01\\
KOI-2462b & 0.060 $\pm$ 0.009\\
KOI-2706b & 0\\
KOI-2801b & 0.00002\\
KOI-3165b & 1\\
KOI-3168b & 0.071 $\pm$ 0.015\\
\end{tabular}
\begin{tablenotes}
 \textbf{Notes.} The FPP is estimated using vespa \citep{morton2016}. With the exception of KOI-1537b and KOI-3165b, which are further excluded from our sample, the false positive probalities are low.
 \end{tablenotes}
\end{center}
\end{table}

\subsection{Eccentricity modeling}
\label{sec:methods_modeling}

The eccentricity of the planet candidate systems is analyzed following the procedure by \cite{vaneylen2015}, with the key aspects of the analysis method summarized here. \textit{Kepler} short cadence observations are reduced starting from the Presearch Data Conditioning (PDC) data \citep{smith2012}, and the planetary orbital period is determined together with any potential transit timing variations (TTVs). The latter is important, as not taking into account TTVs has the potential to bias the eccentricity results \citep[see][for a detailed analysis of the influence of TTVs]{vaneylen2015}. We correct for dilution due to nearby stars following \cite{furlan2017}, who look for nearby (within 4$\arcsec$) companion stars using high-resolution images compiled from other surveys \citep[e.g.][]{howell2011,adams2012,dressing2014,law2014,baranec2016,ziegler2017}. For most systems, the flux contribution of these nearby stars is negligible, but for six systems the `planet radius correction factor' derived by \cite{furlan2017} is larger than 1\%. These systems (and their planet radius correction factor) are KOI-42 (Kepler-410, 3.5\%), KOI-98 (31.7\%), KOI-1962 (33.2\%), KOI-288 (4.4\%), KOI-1537 (38\%), and KOI-1613 (14.1\%). Finally, in six systems, we find TTVs, which are listed in Table~\ref{tab:eccsingle_ttvs}.

\begin{figure}[!htb]
\centering
\resizebox{\hsize}{!}{\includegraphics{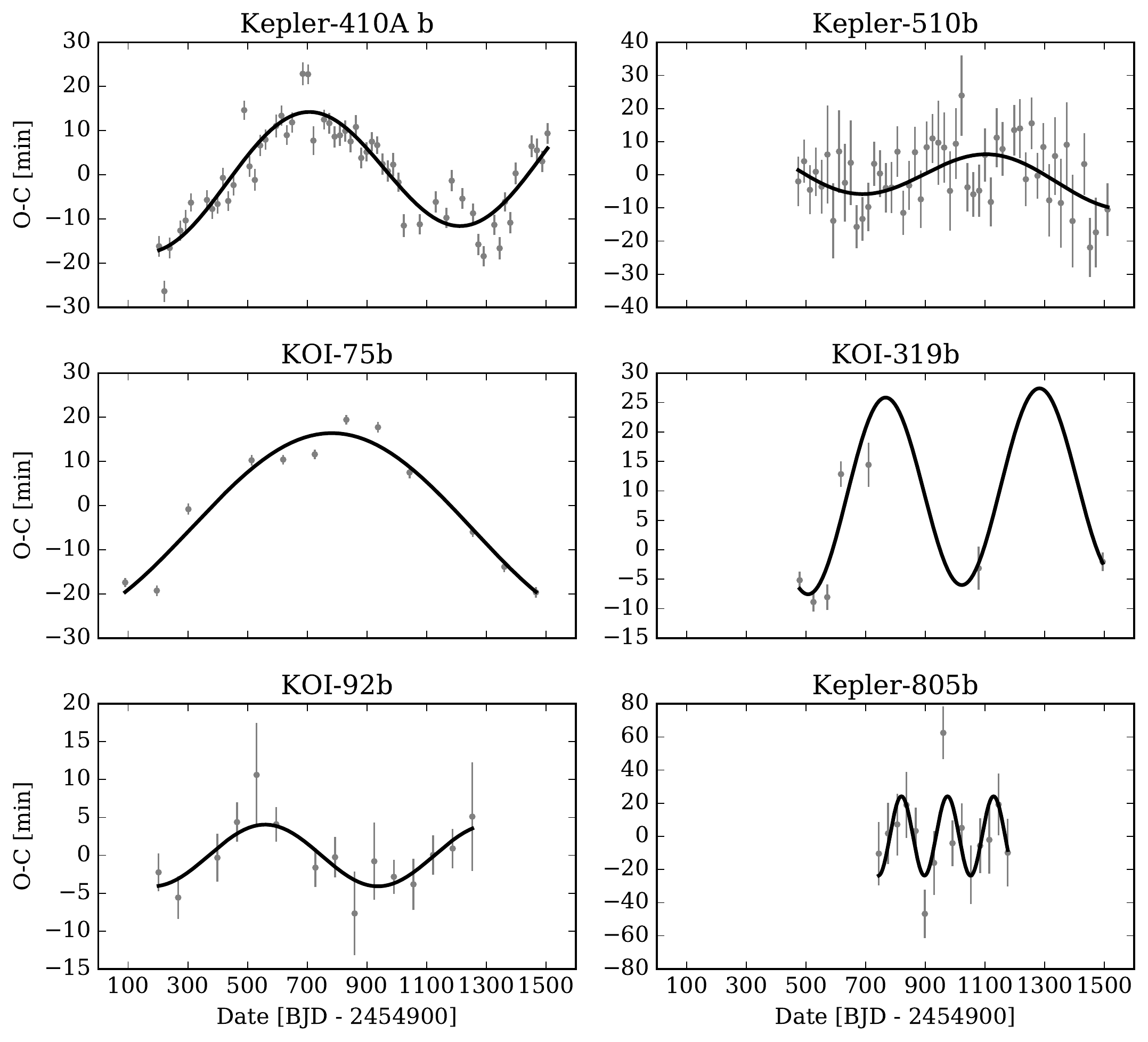}}
\caption{The observed minus calculated transit times are shown for systems with detected TTVs. We fit a sinusoidal model to the O-C times.\label{fig:eccsingle_ttvs}
}
\end{figure}

The data are phase folded, making use of the determined period and a sinusoidal TTV model when TTVs are detected (see Figure~\ref{fig:eccsingle_ttvs}). The transits are then modeled using a Markov Chain Monte Carlo (MCMC) algorithm, specifically an Affine-Invariant Ensemble Sampler \citep{goodman2010} implemented in Python as $emcee$ \citep{foremanmackey2013}. We use the analytical equations by \cite{mandel2002} to model the transit light curves. Eight parameters are sampled, i.e.\ the impact parameter ($b$), planetary relative to stellar radius ($R_\textrm{p}/R_\star$), two combinations of eccentricity $e$ and angle of periastron $\omega$ ($\sqrt{e}\cos \omega$ and $\sqrt{e} \sin \omega$), the mid-transit time ($T_0$), the flux offset which sets the normalization ($F$), and two stellar limb darkening parameters ($u_1$ and $u_2$).

\begin{table}[ht]\begin{center} 
\caption{Overview of the period and amplitude of sinusoidal transit timing variations. The transit times and the best model fits are shown in Figure~\ref{fig:eccsingle_ttvs}. \label{tab:eccsingle_ttvs}}
\footnotesize
\begin{tabular}{lcccccccc}
 	& TTV period [d]	& TTV amplitude [min]\\
 \hline \hline
Kepler-410A~b	& 1055		&	14.4\\
KOI-75b		&	1892	&	21.7\\
KOI-92b		&	756	&	4.1\\
Kepler-510b		&	884	&	7.1\\
KOI-319b		&	515	&	16.3\\
Kepler-805b	&	154	&	23.9\\
\end{tabular}
\end{center}
\end{table}

All parameters are sampled uniformly, i.e.\ using a flat prior, with the exception of the limb darkening coefficients. For the limb darkening coefficients, we used a Gaussian prior centered at values predicted from the table of \citet{claret2011} with a standard deviation of 0.1. We sample in $\sqrt{e} \cos \omega$ and $\sqrt{e} \sin \omega$ rather than in $e$ and $\omega$ directly to avoid a bias due to the boundary condition at zero eccentricity \citep[see e.g.][]{lucy1971,eastman2013}.

Although we use the \cite{mandel2002} equations to model the transits, it is conceptually useful to refer to an approximate equation for the transit duration \citep[see e.g.][]{winn2010},
\begin{equation}
 T = \left( \frac{3}{\pi^2 G} \frac{P}{\rho_\star} \left( 1-b^2 \right)^{3/2} \frac{\left( 1-e^2 \right)^{3/2}}{\left(1+e\sin\omega \right)^3} \right)^{1/3},
\label{eq:transitduration}
 \end{equation}
which is valid for $R_\mathrm{p} \ll R_\star \ll a$. Here, $T$ is the time between the halfway points of ingress and egress, $G$ the gravitational constant, $P$ the orbital period, $\rho_\star$ the mean stellar density, and $a$ the semi-major axis. The final factor in Equation~\ref{eq:transitduration} is sometimes referred to as the density ratio, referring to the ratio between the host star's true density and the `density' derived from the light curve assuming a circular orbit, although the latter is not physically a stellar density. We prefer to refer to the \textit{duration ratio},
\begin{equation}
\label{eq:velocityratio}
\frac{T}{T_\mathrm{circ}} = \frac{\sqrt{1-e^2}}{1+e\sin \omega}.
\end{equation}
Here, $T$ is the measured transit duration (as above), and $T_{\rm circ}$ is the calculated transit duration of a planet on a circular orbit with the same host star, orbital period, and impact parameter.  Equation~\ref{eq:transitduration} shows that to calculate $T_{\rm circ}$ we need to know the period, impact parameter, and mean
stellar density.  The orbital period is known precisely from the measured times of individual transits.
The mean stellar densities in our sample come from asteroseismology (see Section~\ref{sec:sampleselection}).
The impact parameter can be derived by fitting the transit light curve, although the uncertainty in the impact
parameter can be strongly covariant with that of the transit duration -- hence the importance of the MCMC modeling
procedure described above, in which all the parameters and their covariances are determined.

Measurement of the impact parameter is prone to observational biases \citep{vaneylen2015}, e.g.\ dilution due to nearby stars, or TTVs, which is why these effects are taken into account, as described above. In addition to correcting for TTVs ourselves, we also cross-check our sample with the TTV catalogue by \cite{holczer2016}. For KOI-42 (Kepler-410), KOI-75, and KOI-319, the TTVs we measured were also detected by \cite{holczer2016}. For KOI-374, \cite{holczer2016} find TTVs with a very long orbital period (1388 days). However, only four of the transits were observed in short-cadence and they are best fitted with a linear trend, which is absorbed into the calculated orbital period. We find evidence of TTVs in KOI-92, KOI-281 (Kepler-510), and KOI-1282 (Kepler-805; see Figure~\ref{fig:eccsingle_ttvs}), which were not detected by \cite{holczer2016}. As a robustness check, we also modeled these systems without including TTVs and found no significant effect on the derived eccentricities.

\begin{figure}[!htb]
\centering
\resizebox{\hsize}{!}{\includegraphics{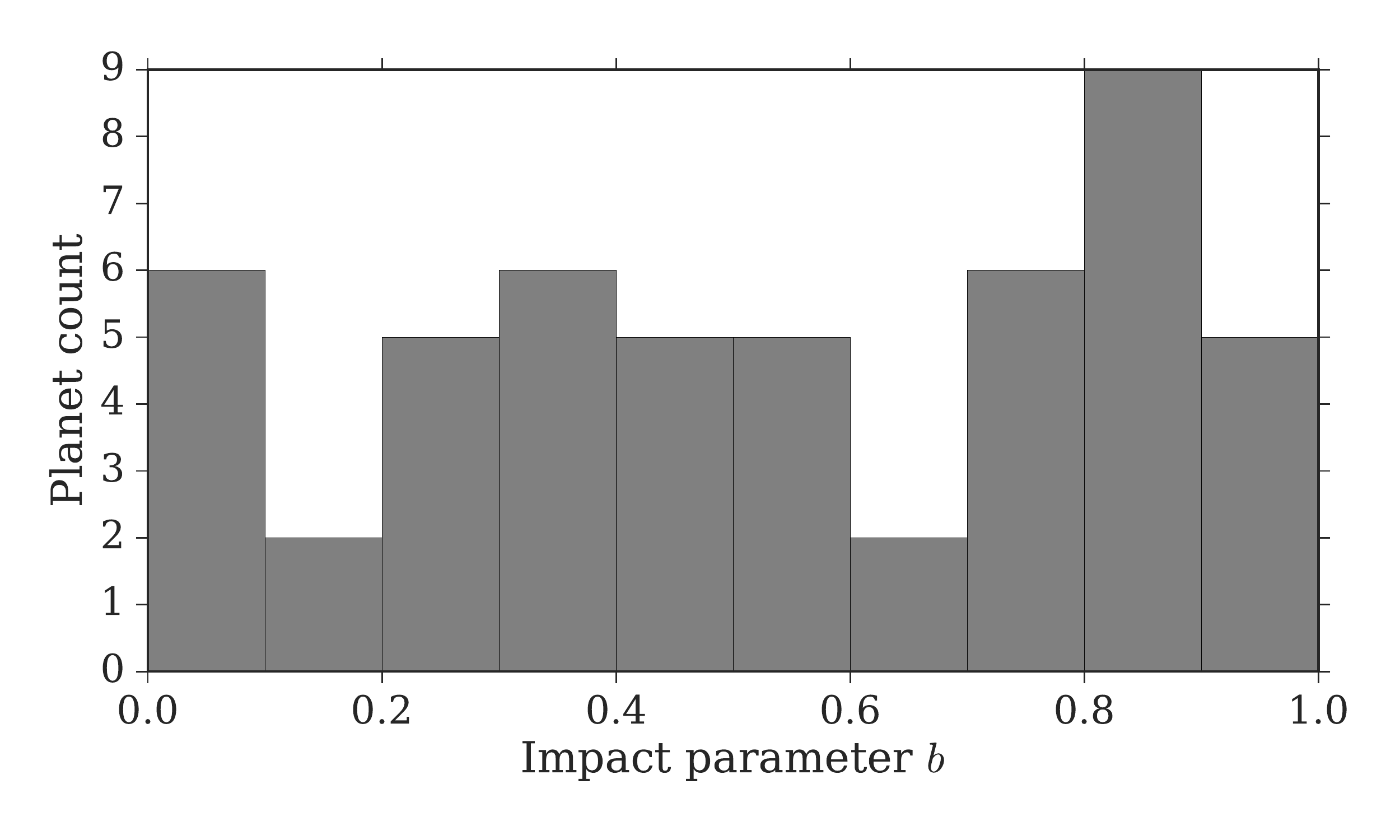}}
\resizebox{\hsize}{!}{\includegraphics{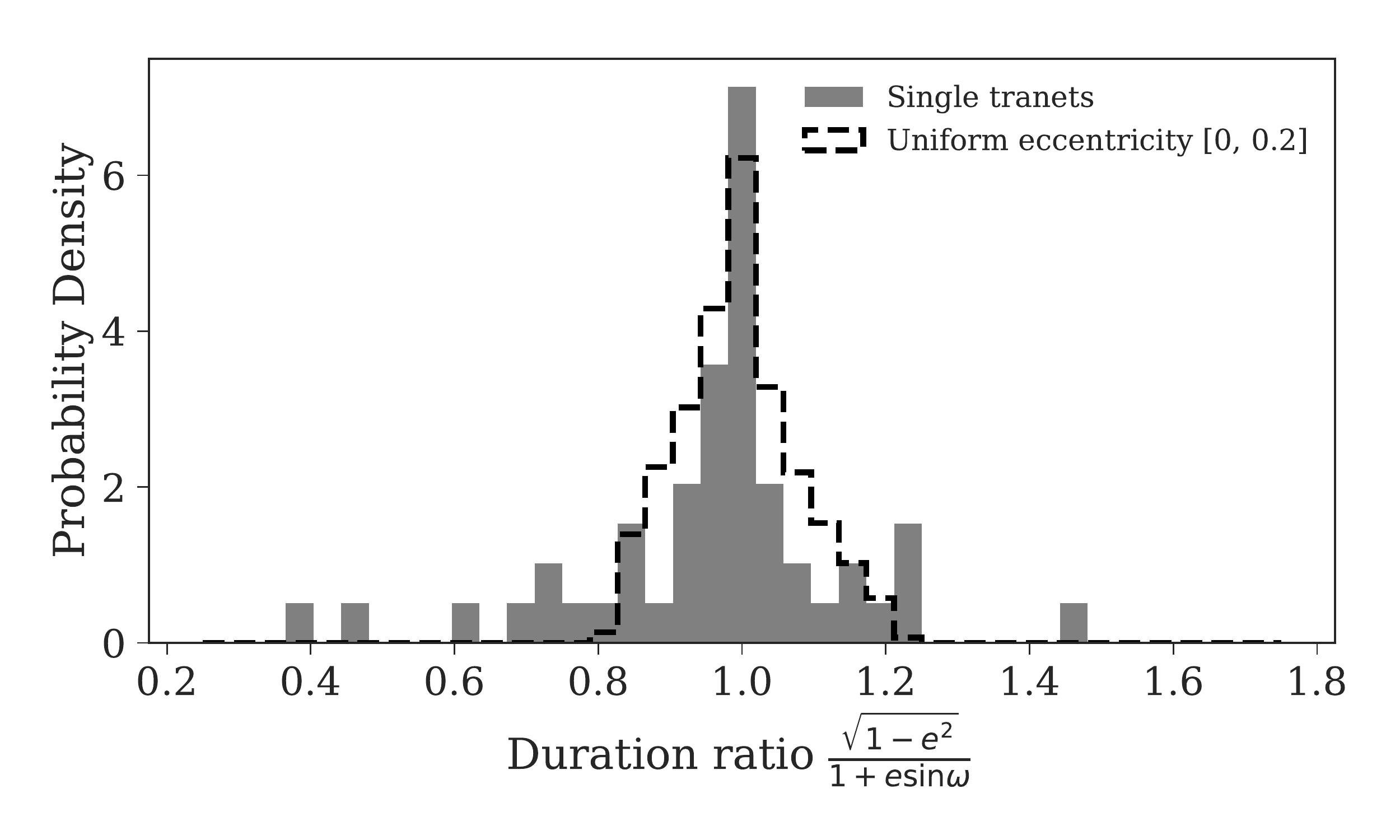}}
\caption{\textit{Top:} histogram showing the modal values for the impact parameter for the planets in our sample. Impact parameters should be distributed approximately uniformly, which serves as a check for the modeling. \textit{Bottom:} histogram showing the duration ratio, which is a combination of eccentricity and angle of periastron that transits can constrain. As an illustration, we overplot the duration ratio distribution for a uniform distribution in eccentricity between 0 and 0.2, with randomly chosen angles of periastron (corrected for the transit probability). \label{fig:eccsingle_impactparams}
}
\end{figure}

To check for any bias affecting the impact parameter measurements, we can check its distribution.  For a random sample of transiting planets, the distribution of impact parameters is expected to be approximately flat between zero and one, although higher impact parameter values are suppressed because they correspond to a lower signal-to-noise ratio \citep{kipping2016}. A histogram of the modes of the impact parameters for the planets in our sample is shown in Figure~\ref{fig:eccsingle_impactparams}, and they appear roughly uniformly distributed.
To quantify this, we run a Kolmogorov-Smirnov test comparing the best values for the impact parameters to a uniform distribution, and find a test statistic of $0.11$ and a $p$-value of 0.57, which indicates that we cannot reject the null hypothesis that impact parameters are uniformly distributed between 0 and 1.

As a final consistency check, we can look at the distribution of the duration ratio (Equation~\ref{eq:velocityratio}). For circular orbits, the duration ratio is always one. For low eccentricities, a distribution around unity is expected. In Figure~\ref{fig:eccsingle_impactparams}, we plot the distribution of the duration ratio for a uniform distribution of eccentricity between 0 and 0.2, with uniform angles of periastron (assuming we do not occupy a special place in the universe) corrected for the transit probability which is proportional to $(1+e\sin\omega)/(1-e^2)$ \citep[e.g.][]{barnes2007,kipping2014_occurrencerate}. Here, we find a peak in the duration ratio distribution at unity, with values distributed roughly evenly on both sides, i.e.\ 52\% are below unity and 48\% are above. We overplot the observed duration ratios in Figure~\ref{fig:eccsingle_impactparams}, and find the same features, i.e.\ a peak at unity, 28 duration ratios below unity and 23 above, revealing no obvious biases. The eccentricity distribution is further analyzed in Section~\ref{sec:distributionresult}.

\section{Results}
\label{sec:results}

\subsection{Individual systems} 

Table~\ref{tab:eccsingle_paramtable} gives the eccentricity constraints for the 51 single-tranet systems in our sample. Some individual systems are briefly discussed in Appendix~\ref{sec:appendix_individual}. 
Figure~\ref{fig:periodecc_sample} gives an overview of the eccentricity measurements as a function of orbital period, with the symbol size proportional to planet radius. Planet candidates and confirmed planets are distinguished with different symbols.

At short orbital periods (e.g.\ $P < 5$~days), most planets show low orbital eccentricities, consistent with zero, as expected due to tidal circularization. There are a few exceptions. HAT-P-11b shows a moderate eccentricity, which is also observed with radial velocity measurements \citep{bakos2010}, and Kepler-21b may also exhibit a moderate eccentricity. At face value, Kepler-408b shows a significant eccentricity, but caution is warranted as its orbit is consistent with zero eccentricity within 95\% confidence. 

\begin{figure*}
\centering
\resizebox{\hsize}{!}{\includegraphics{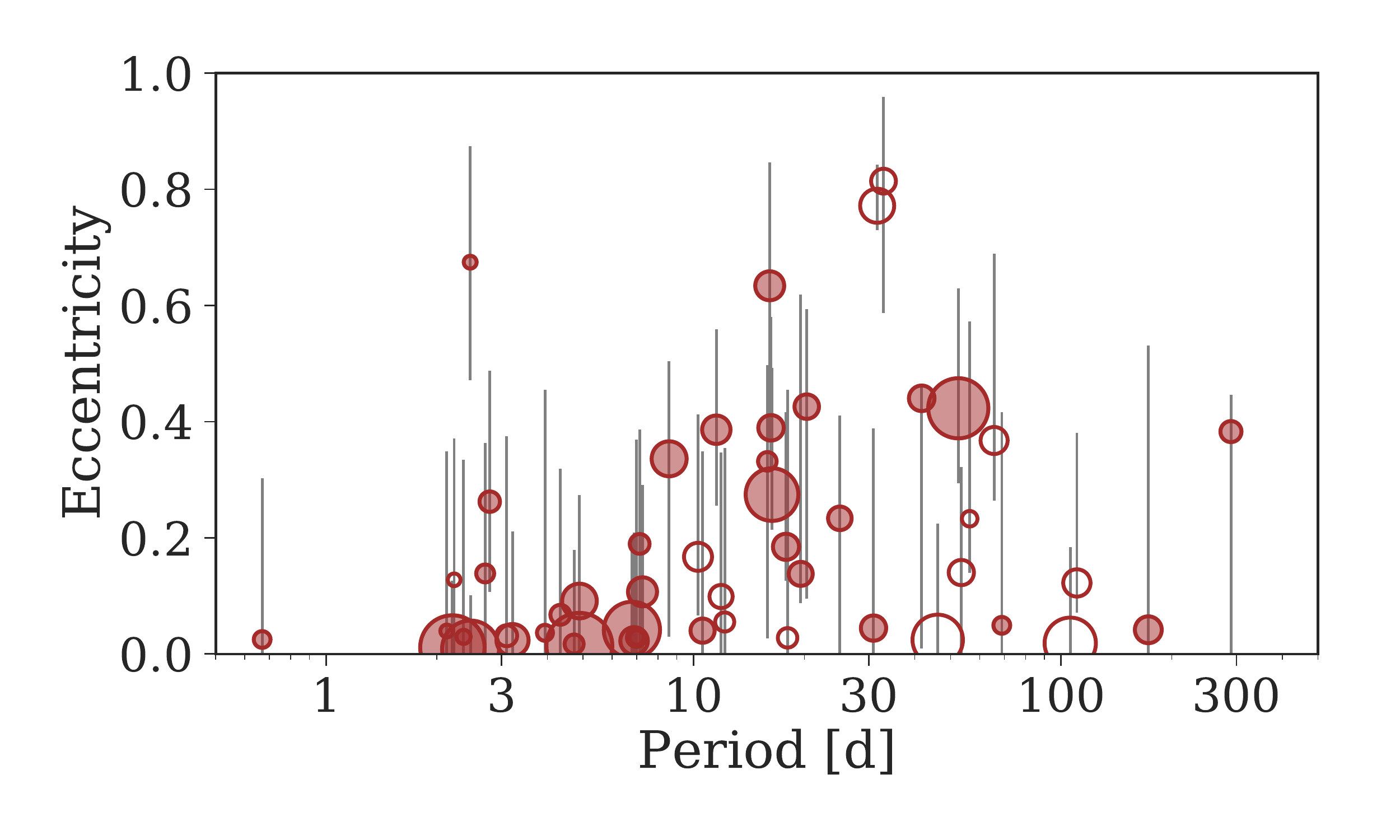}}
\caption{Overview of our sample. The measured orbital eccentricities are plotted as a function of orbital period. The symbol size is proportional to the planet radius. Open circles represent planet candidates, while filled circles represent confirmed planets.\label{fig:periodecc_sample}
}
\end{figure*}

\begin{figure*}
\centering
\resizebox{\hsize}{!}{\includegraphics{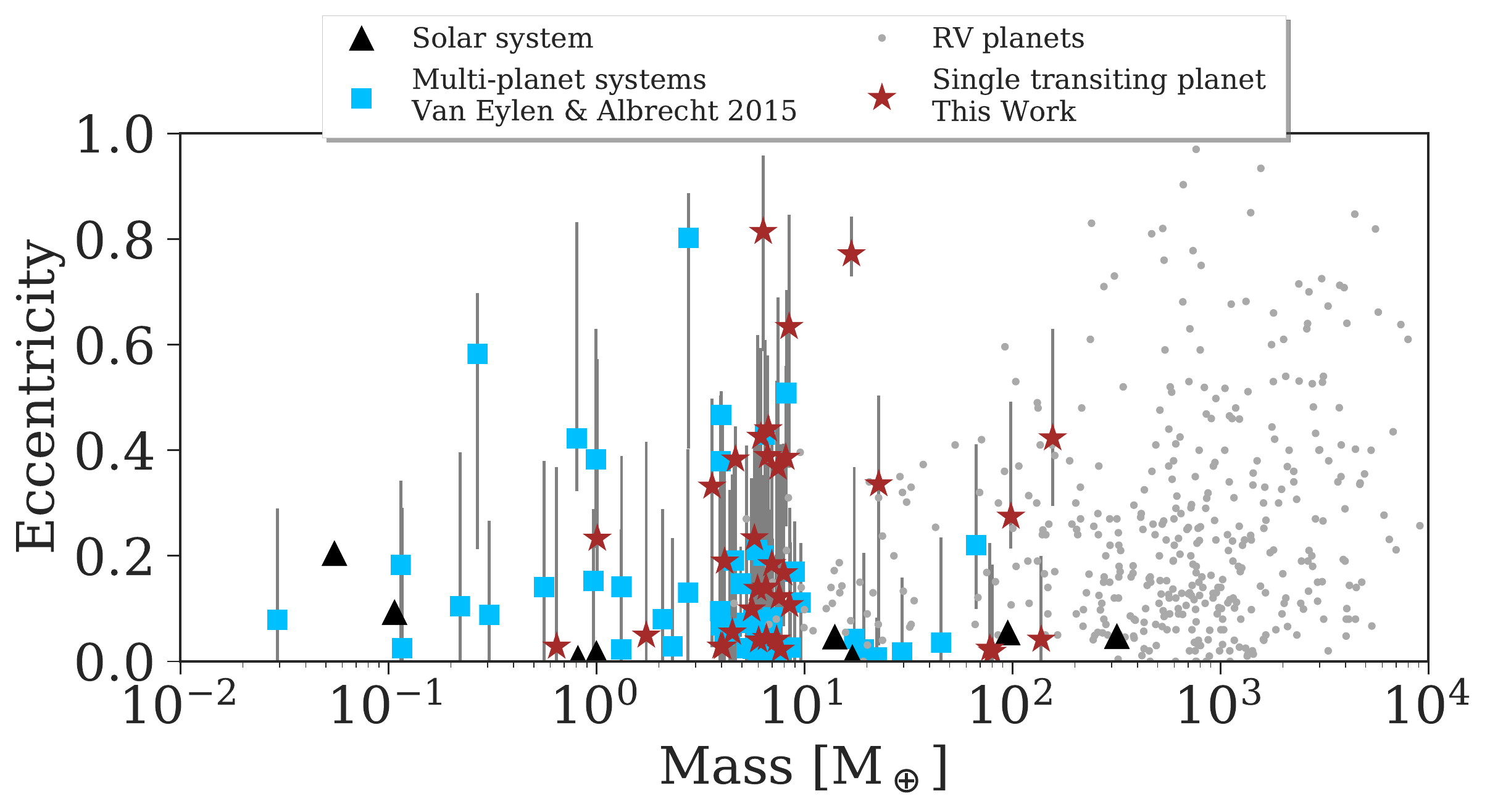}}
\caption{The eccentricity and mass measurements for exoplanets are plotted as taken from exoplanets.org on 8 December 2017, for planets where both values are determined through radial velocity. For comparison, the solar system is shown. The eccentricities of multi-planet systems determined by \cite{vaneylen2015} are shown in blue and the eccentricities of the planets in our sample are plotted in red, with the planet masses estimated based on radius \citep{weiss2013,weiss2014}. In all cases, only planets with orbital periods longer than five days are plotted. \label{fig:masseccentricity}
}
\end{figure*}

In Figure~\ref{fig:masseccentricity}, we show our eccentricity measurements together with eccentricity measurements determined using radial velocities, as well as with the eccentricities of transiting multi-planet systems \citep{vaneylen2015}. We estimate the masses of the transiting planets from the radius, following \cite{weiss2013} and \cite{weiss2014}. We show only planets with orbital periods longer than five days, to avoid being dominated by tidally circularized systems.

\subsection{Eccentricity distribution} 
\label{sec:distributionresult}

\subsubsection{Hierarchical inference procedure}
\label{sec:dfmmethod}

We now determine the overall distribution of orbital eccentricities. 
To do so, we employ a hierarchical inference procedure outlined by \cite{hogg2010} and further developed by \citet{foremanmackey2014} (see Section 3 therein). In this method, we directly use the posterior distribution of eccentricities determined from our MCMC fitting procedure (see Section~\ref{sec:methods_modeling}) to individual planets to infer the distribution of eccentricities for a sample (or subsample) of planets. 
The likelihood function of the observed set of eccentricities for all individual planets, given a distribution of eccentricities for the sample described by parameters $\theta$, assuming that the eccentricity of planets orbiting different stars is independent, is given by \citep{foremanmackey2014}

\begin{equation}
 p (\mathrm{obs} | \theta) \propto \frac{1}{N} \prod_{k=1}^{K} \sum_{n=1}^{N} \frac{p(e_k^{n} | \theta)}{p(e_k^n | \alpha)}.
\end{equation}

Here, $p(e_k^{n} | \theta)$ is the probability density of a certain eccentricity ($e$) given the model with parameters $\theta$ -- we will proceed to try several different models, such as a Gaussian and a Beta distribution. $p(e_k^n | \alpha)$ is the prior probability of this value. In our case, this is simply a constant, as we assumed uniform priors for the eccentricity (see Section~\ref{sec:methods_modeling}). These values are multiplied over $K$ different exoplanets, and summed over $N$ different posterior values for each planet. 

We then determine the parameters $\theta$ of the eccentricity distribution by optimizing the likelihood $p (\textrm{obs} | \theta)$. We multiply by a uniform prior on the parameters $\theta$ and use the MCMC algorithm with affine-invariant sampling, \textit{emcee} \citep[][]{foremanmackey2013}, to sample the posterior, and use 10 walkers each carrying out 10,000 steps, including a burn-in phase of 5,000 steps. We report median values and 68\% highest probability density limits. In all cases, we impose a uniform prior on the distribution parameters. For the posterior distribution of the individual planets, we randomly select 100 posterior samples for each planet, to ensure computational tractibility.

\subsubsection{Simple eccentricity distribution}
\label{sec:distributionfit}

We use the mechanism described in Section~\ref{sec:dfmmethod} to model the empirical distribution of eccentricities, which we later compare to simulations (see Section~\ref{sec:discussion}). Because the eccentricity excitation mechanisms may be different for Earth and super-Earth systems than for systems with transiting Jupiters, from here on, we limit the sample to small planets ($R < 6~R_\oplus$), and to avoid the possible influence of tides we further only take long orbital periods ($P>5$~days). We look into short-period planets in Section~\ref{sec:shortperiodplanets}, where we also test the sensitivity to the cut-off period, and investigate giant planets in Section~\ref{sec:giantplanets}.

We try to fit several different distributions to the observed eccentricities. We fit these to the single-tranet systems measured here, and separately to the multi-tranet systems in \cite{vaneylen2015}. As a simple case, we try a Rayleigh distribution, which has a single parameter $\sigma$. We find $\sigma = 0.24^{+0.04}_{-0.04}$ for the single-tranets, and $\sigma = 0.061^{+0.010}_{-0.012}$ for the multi-tranets. 
The latter is comparable to $\sigma = 0.049 \pm 0.013$, as determined by \cite{vaneylen2015}. However, unlike the procedure described in Section~\ref{sec:dfmmethod}, the fitting procedure in \cite{vaneylen2015} used only best-fit values and did not take into account the posterior distributions of the eccentricity observations.

With eccentricities close to zero, we also try a half-Gaussian distribution, i.e.\ the positive half of a Gaussian distribution that peaks at zero. We fit for the width of the distribution, and find $\sigma = 0.32^{+0.06}_{-0.06}$ and $\sigma = 0.083^{+0.015}_{-0.020}$ for single-tranet and multi-tranet systems, respectively. 

The use of a Beta distribution has also been advocated in the context of orbital eccentricities \citep[e.g.][]{kipping2013}, which has the advantage of mathematically flexible properties. This makes it suitable to look for differences in the underlying distribution, without knowing its exact shape, and also convenient to use as a prior in (future) transit fits. Such a distribution has two parameters, i.e.\ $\theta = \{a,b\}$. For single-tranet systems, we find $\{1.58^{+0.59}_{-0.93}, 4.4^{+1.8}_{-2.2}\}$, while for systems with multiple transiting planets we find $\{1.52^{+0.50}_{-0.85}, 29^{+9}_{-17}\}$. 

We fit a mixture model which contains a half-Gaussian and a Rayleigh distribution, where the former captures low-eccentricity systems, while the latter can encapsulate higher-$e$ planets. We are fitting for $\theta = \{\sigma_\mathrm{Gauss}, \sigma_\mathrm{Rayleigh}, f_\mathrm{single}, f_\mathrm{multi}\}$, i.e.\ the width of the half-Gaussian, the Rayleigh parameter, and the fraction of the Gauss and Rayleigh components, for single- and multi-tranet systems respectively. Here, $f=0$ indicates a pure half-Gaussian distribution, and $f=1$ a pure Rayleigh distribution. 

We find $\theta = \{0.049^{+0.017}_{-0.024}, 0.26^{+0.04}_{-0.06}, 0.76^{+0.21}_{-0.12}, 0.08^{+0.03}_{-0.08}\}$. These results are consistent with the simple distributions derived above: the majority of single-tranet systems have a significant eccentricity, while almost all multi-tranet systems have low eccentricities. 

\begin{figure*}[!htb]
\centering
\resizebox{0.49\hsize}{!}{\includegraphics{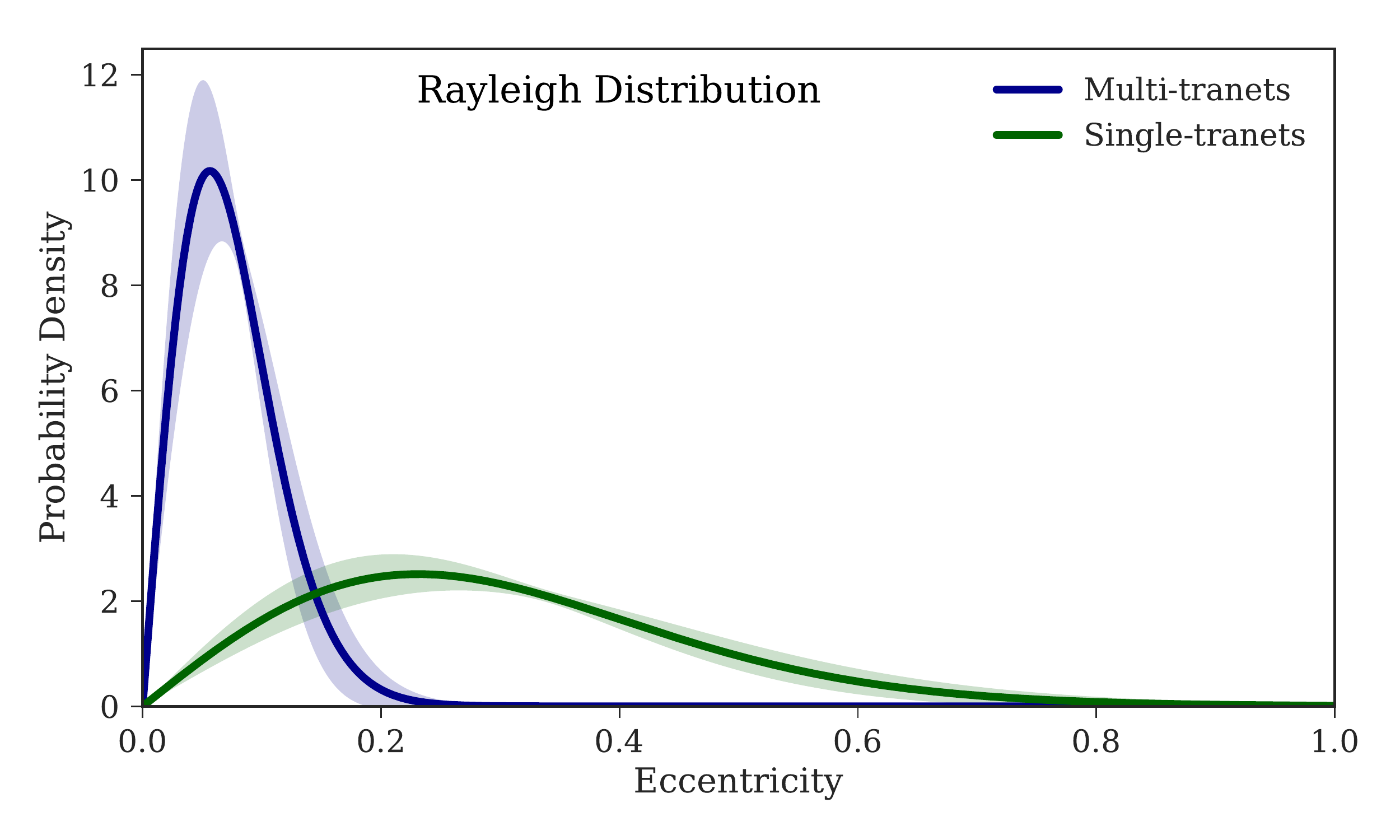}}
\resizebox{0.49\hsize}{!}{\includegraphics{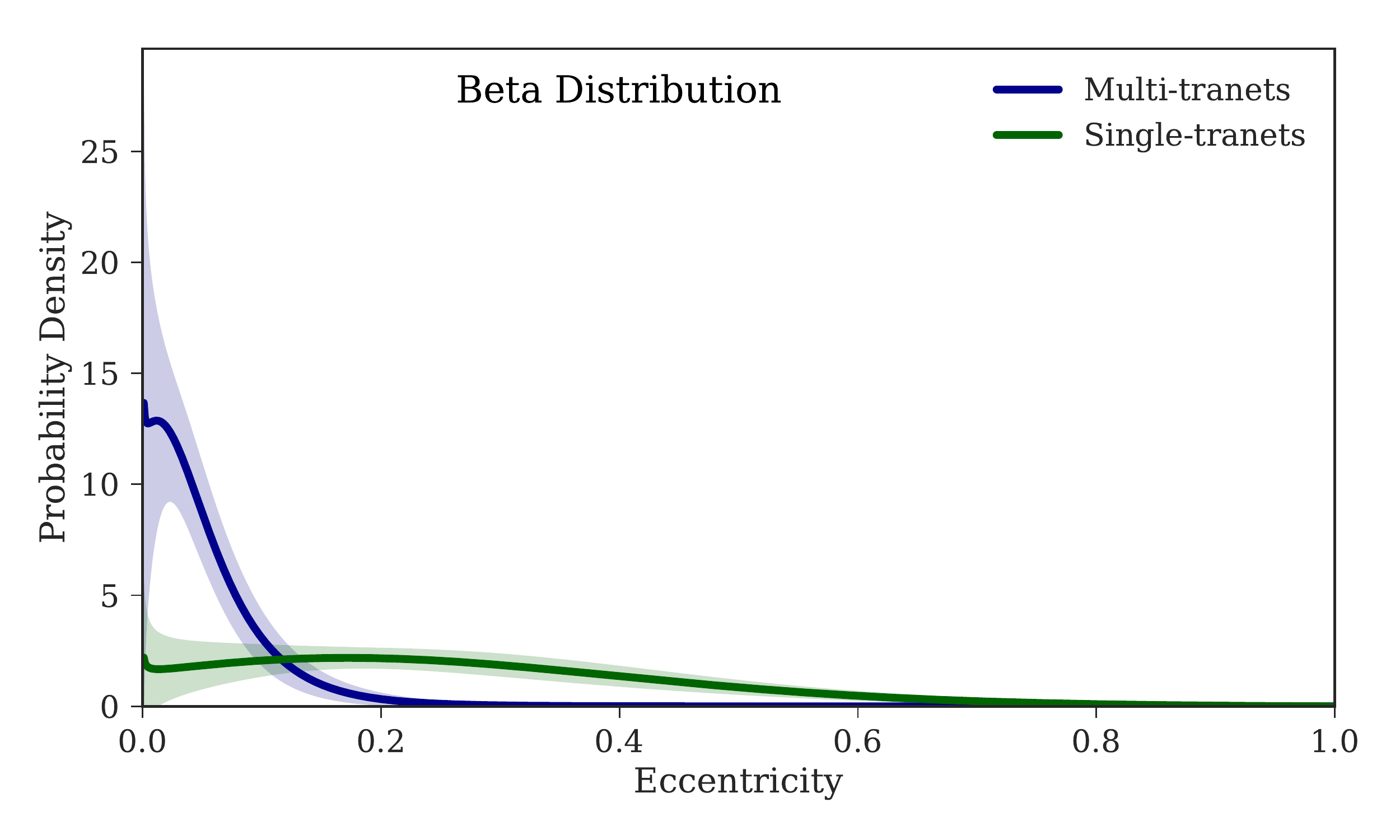}}
\resizebox{0.49\hsize}{!}{\includegraphics{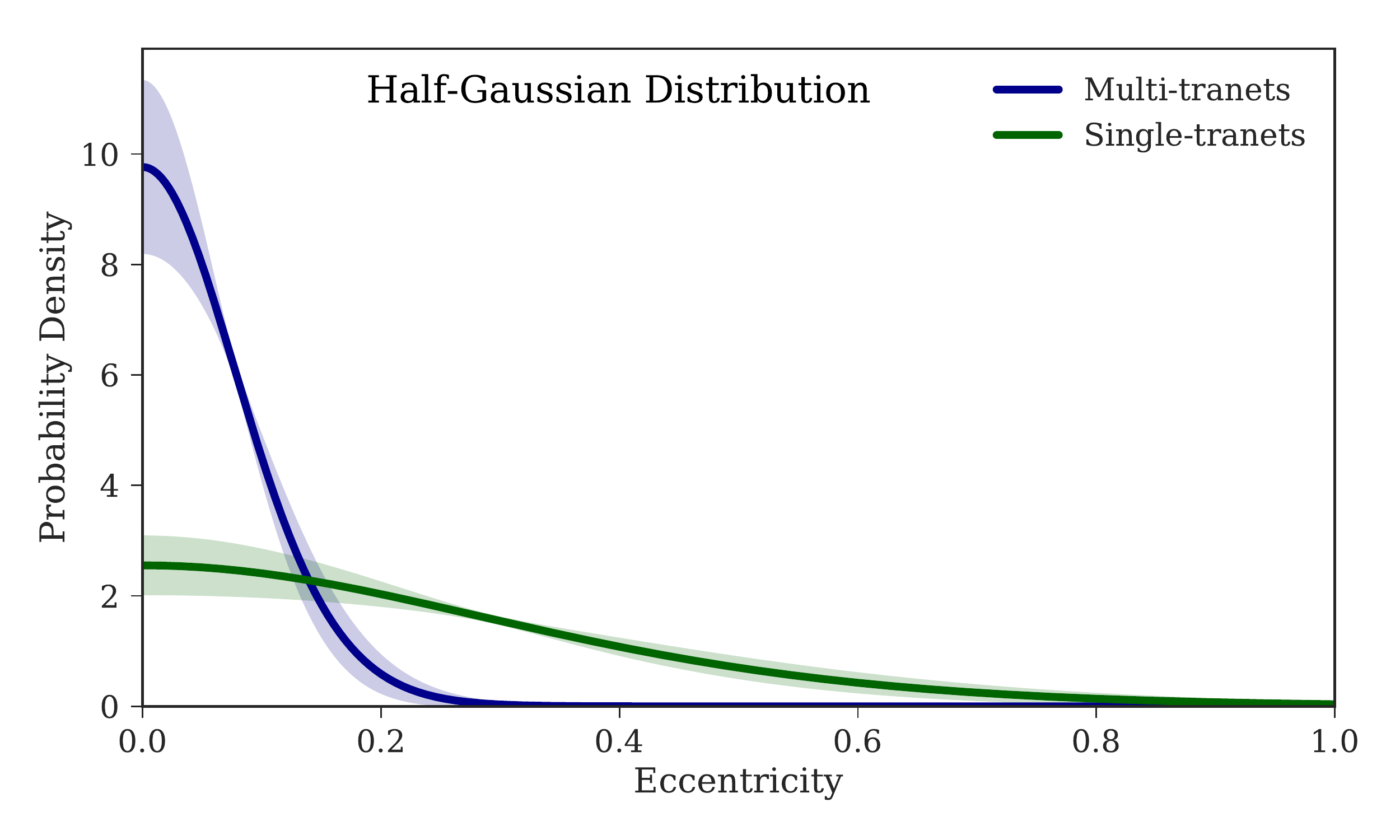}}
\resizebox{0.49\hsize}{!}{\includegraphics{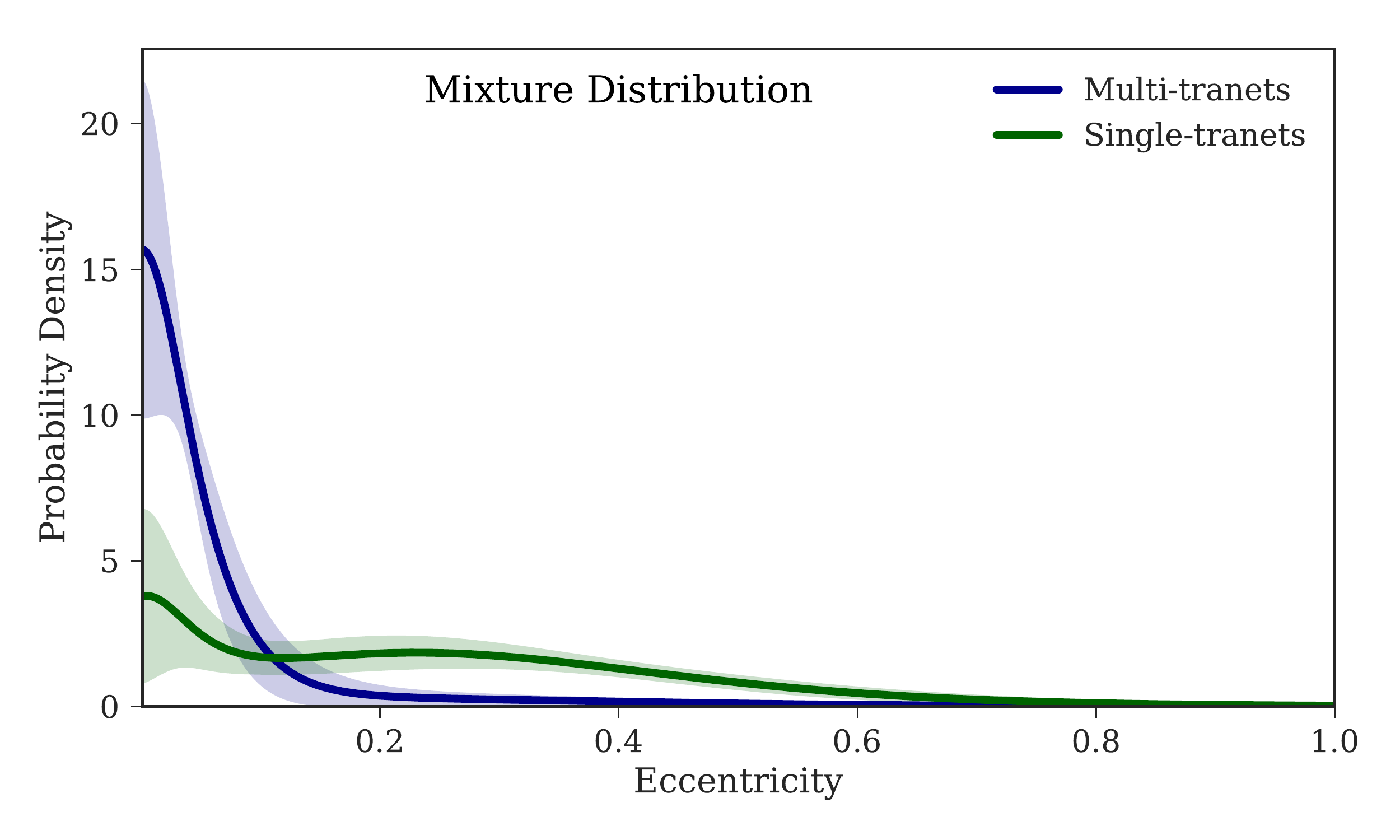}}
\caption{The best fit distributions to the multi- and single-tranet systems. The colored area represents a 68\% confidence interval of the distribution, and the thick line shows the median value. \textit{Top left:} a Rayleigh distribution. \textit{Bottom left:} a half-Gaussian distribution. \textit{Top right:} a Beta distribution. \textit{Bottom right:} mixture model with a half-Gaussian and a Rayleigh component.\label{fig:distributionfits}
}
\end{figure*}

We show all these distributions in Figure~\ref{fig:distributionfits} and summarize their parameters in Table~\ref{tab:distributions}. We do not advocate for one model over another, and report the parameters for all the models we have tested. In all these cases, we find a clear difference in the eccentricity distribution between single- and multi-tranet systems.

\begin{figure}[!htb]
\centering
\resizebox{\hsize}{!}{\includegraphics{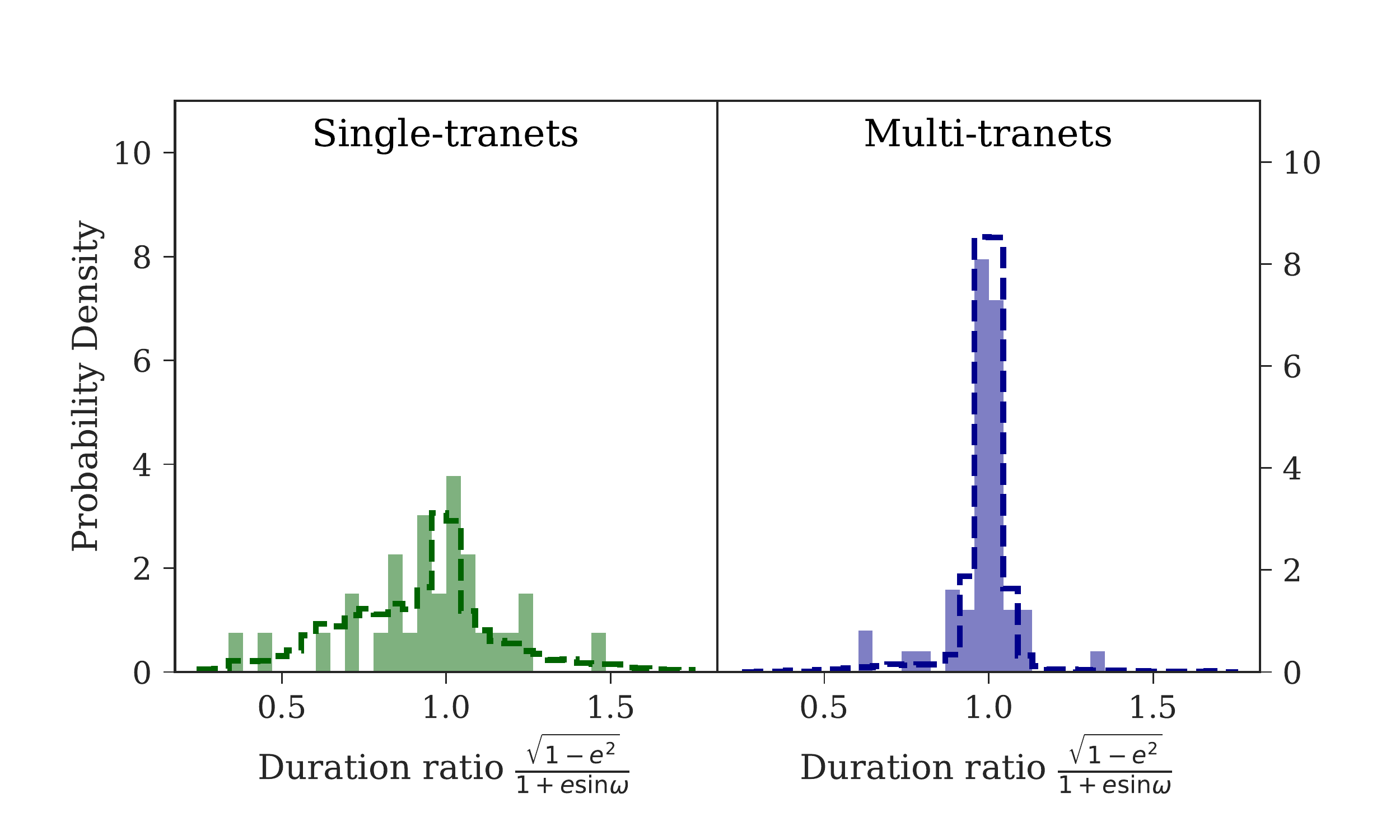}}
\caption{Comparison of the duration ratio for single-tranet and multi-tranet systems. The colored histogram bars show the median values for each planet, for $P>5~$days and $R<6~R_\oplus$. The dashed line is calculated based on the best fitting mixture model, with uniformly distributed angles of periastron, but after correcting for the transit probability. Note that this is not a direct fit, as the fitting procedure includes the full posterior distributions rather than best values. \label{fig:velrhobestfit}
}
\end{figure}

\begin{figure}[!htb]
\centering
\resizebox{\hsize}{!}{\includegraphics{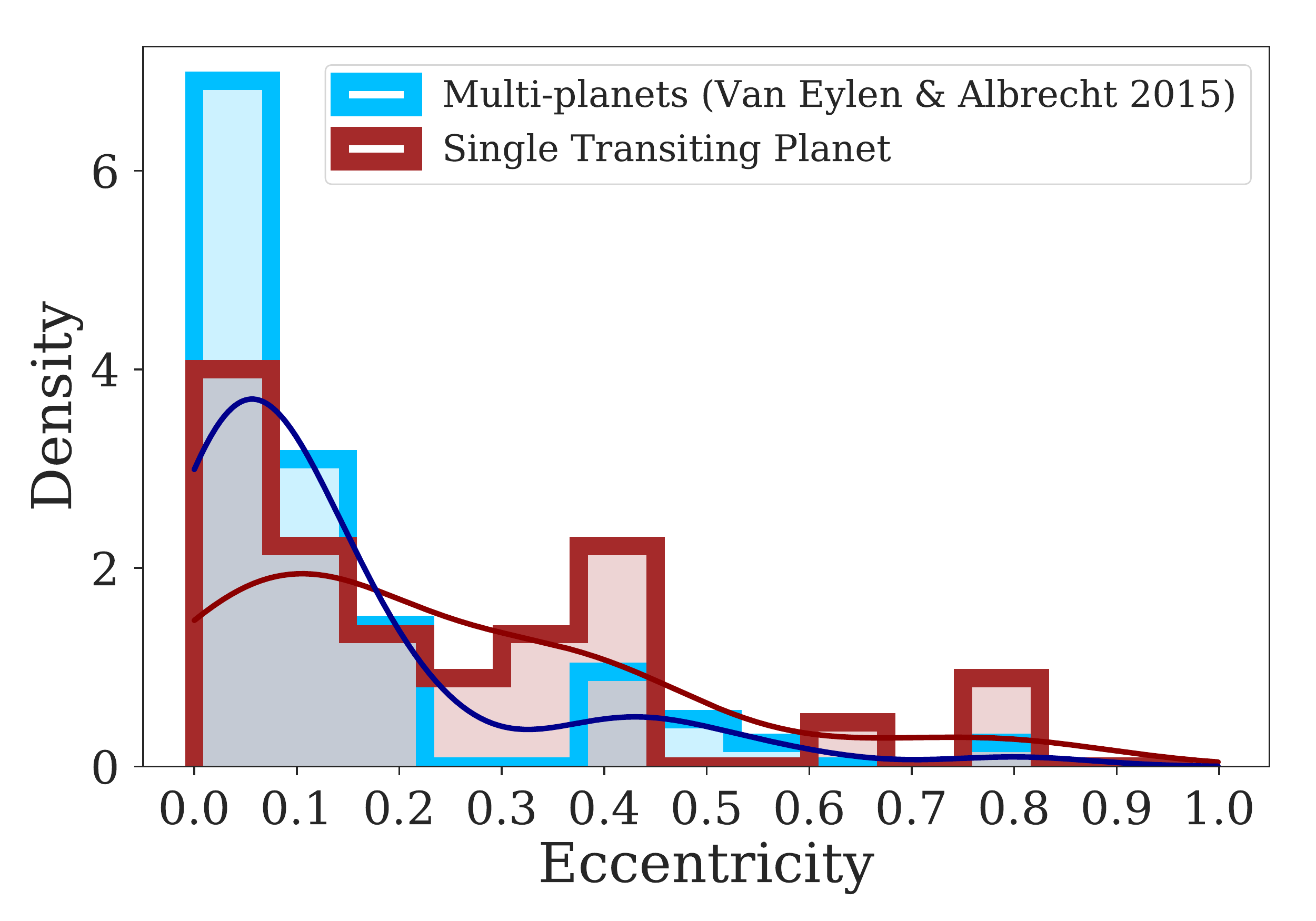}}
\caption{Histogram showing the eccentricity distribution for systems with multiple transiting planets and systems with a single transiting planet. The bins have an arbitrarily chosen width of 0.075. To mitigate the influence of the bin width, we also show a kernel density estimate (colored lines), where the width was determined using Scott's rule. Multi-planet systems clearly have a higher density of low eccentricities. This histogram only uses best values for illustration, to determine the eccentricity distribution we use the full posterior distribution instead (see Section~\ref{sec:dfmmethod}). \label{fig:histogram_eccentricity}
}
\end{figure}

All these distributions show a similar behavior, except at zero eccentricities. As can be seen from Figure~\ref{fig:distributionfits}, the flexible Beta distribution also allows for a range of probability densities at zero eccentricity.  We show what this distribution looks like in duration ratio space (see Equation~\ref{eq:velocityratio}), by matching the eccentricity distribution to random (uniform) angles of periastron $\omega$, and weighing them by the transit probability. This is shown in Figure~\ref{fig:velrhobestfit}, together with best (median) values for the duration ratio for our sample of planets as well as the multi-tranet systems. Although our fitting procedure is more complex than simply comparing best values (see Section~\ref{sec:dfmmethod}), it is reassuring to see that the fitted distributions match the overall shape of the observed duration ratios.
As a final consistency check, we also plot a simple histogram of eccentricities, simply using best value estimates rather than the full posteriors we adopted in the hierarchical Bayesian procedure here. This is shown in Figure~\ref{fig:histogram_eccentricity}. A clear overdensity of low-eccentricity planets is seen for multi-tranet systems, consistent with the results of the hierarchical Bayesian methods, which use the full posterior distribution of individual systems.

\begin{table*}[htbp]
\begin{center}
\footnotesize
\caption{Eccentricity distributions\label{tab:distributions}}
\begin{tabular}{lcc}
Distribution & Parameters & Best Values \\
 \hline \hline
Rayleigh & $\{ \sigma_\mathrm{single}, \sigma_\mathrm{multi} \}$ & $\{0.24^{+0.04}_{-0.04}, 0.061^{+0.010}_{-0.012} \}$\\[0.2cm]
Half-Gaussian & $\{ \sigma_\mathrm{single}, \sigma_\mathrm{multi} \}$ & $\{0.32^{+0.06}_{-0.06}, 0.083^{+0.015}_{-0.020} \}$\\[0.2cm]
Beta & $\{ a_\mathrm{single}, b_\mathrm{single}, a_\mathrm{multi}, a_\mathrm{multi} \}$ & $\{1.58^{+0.59}_{-0.93}, 4.4^{+1.8}_{-2.2}, 1.52^{+0.50}_{-0.85}, 29^{+9}_{-17} \}$\\[0.2cm]
Mixture & $\{ \sigma_\mathrm{Gauss}, \sigma_\mathrm{Rayleigh}, f_\mathrm{single}, f_\mathrm{multi} \}$ & $ \{0.049^{+0.017}_{-0.024}, 0.26^{+0.04}_{-0.06}, 0.76^{+0.21}_{-0.12}, 0.08^{+0.03}_{-0.08} \}$\\[0.2cm]
\end{tabular}
\begin{tablenotes}
 \textbf{Notes.} Details for the fitting parameters are provided in Section~\ref{sec:distributionfit}. Only planets with $R < 6~R_\oplus$ and $P>5$~days are included. The best values are median values and 68\% highest probability density intervals.
\end{tablenotes}
\end{center}
\end{table*}

\subsubsection{Planet candidates and confirmed planets}

Is the eccentricity distribution influenced by the presence of planet candidates, i.e.\ objects that have not yet been confirmed as bona fide planets? The multi-planet systems by \cite{vaneylen2015} consist nearly exclusively of confirmed planets, but the single-tranet systems observed contain 17 planet candidates out of 53 systems. 

As argued in Section~\ref{sec:sampleselection}, we estimate that this sample contains at most one or two false positives.  This leads us to wonder if the
results are being skewed by one or two outliers.  To test this, we remove the two systems with the highest eccentricity posterior, i.e.\ KOI-367b and KOI-1962b, both unconfirmed planet candidates. For simplicity, we compare the results for a half-Gaussian distribution. For this distribution, we now find $\sigma = 0.24^{+0.05}_{-0.06}$, consistent with the value for the full distribution within 1.5$\sigma$ (see Table~\ref{tab:distributions}).  Thus, the results do not seem to be especially sensitive to the few points with the highest eccentricities.

To further investigate if false positives could influence our result, we exclude all planet candidates from the sample, and only model the confirmed planets. Again comparing a half-Gaussian distribution, we find $\sigma = 0.27^{+0.06}_{-0.08}$ and $\sigma = 0.083^{+0.016}_{-0.020}$, for single- and muli-tranet systems, respectively. Both results are consistent at the $1\sigma$ level with the values found for the full sample (see Table~\ref{tab:distributions}), again suggesting that false positives are not responsible for the observed differences between multis and singles.

\subsubsection{Planet size and orbital period for singles and multis}

We also investigate the role of planet size on orbital eccentricity.  Single-tranet systems and multi-tranet systems may have systematically different planet size distributions (see Figure~\ref{fig:masseccentricity}).  This raises the possibility that the observed difference between single- and multi-tranet systems is a side effect of the different planet size distributions.  To enforce a similar distribution for planet radius, we divide the sample into bins of $1~R_\oplus$, between 1 and 6~$R_\oplus$, and select an equal number of single-tranet and multi-tranet systems in each bin, which leads to a sample of 26 planets in each category, with the same distribution of planet sizes. 

We apply the modeling procedure described above on these new subsamples and for a half-Gaussian distribution, we find $\sigma = 0.34^{+0.07}_{-0.07}$ and $\sigma = 0.060^{+0.019}_{-0.030}$ for single-tranet and multi-tranet systems, respectively. These results are consistent at about $1\sigma$ with the distributions determined for the full sample, as listed in Table~\ref{tab:distributions}. The difference between single- and multi-tranet systems remains the same.  We conclude that differences in planet size are not likely to be responsible for the different eccentricity distributions inferred for single- and multi-tranet systems.

The orbital period distributions of single- and multi-tranet systems are very similar, but we nevertheless apply the same procedure to test the influence of this parameter. We divide the sample into bins with a width of 0.5 in the logarithm (base 10) of the orbital period, between 0.5 and 2.5.  We select an equal number of single- and multi-tranet systems in each bin, to generate a subsample of 29 single- and 29 multi-tranet systems. Once again, comparing a half-Gaussian distribution, we find $\sigma = 0.33^{+0.05}_{-0.07}$ and $\sigma = 0.096^{+0.022}_{-0.027}$ for single-tranet and multi-tranet systems, respectively. These values are consistent within $1-2\sigma$ with the distribution determined for the full sample (see again Table~\ref{tab:distributions}).

\subsubsection{Short-period planets}
\label{sec:shortperiodplanets}

So far, we have limited our sample to systems with $P>5$~days. We excluded the short-period systems because they are likely to have been influenced by tidal circularization. 
We now model the systems with $P<5$~days (while still retaining the same upper limit on the radius of $6~R_\oplus$). There are 13 such single-tranet systems and 6 such multi-tranets. For a half-Gaussian distribution, we find $\sigma_\mathrm{single} = 0.10^{+0.03}_{-0.05}$ and $\sigma_\mathrm{multi} = 0.04^{+0.03}_{-0.04}$. As expected, the orbital eccentricity peaks closer to zero, validating our choice to exclude these systems from our previous analysis. It appears that single-tranet short-period systems may be slightly more eccentric than multi-tranet short period systems, but the distributions are consistent at the $1\sigma$ level. 

To check whether a period of $5$~days is a sensible cut, we check what happens to the overall eccentricity distribution if we only include planets at $P>10$~days. For a half-Gaussian distribution, we now find $\sigma_\mathrm{single} = 0.36^{+0.06}_{-0.08}$ and $\sigma_\mathrm{multi} = 0.09^{+0.01}_{-0.04}$. These values are slightly larger (i.e.\ more eccentric) than when a cut-off at $P=5$~days is used (see Table~\ref{tab:distributions}), but consistent at the $1\sigma$ level. Similarly, if we use a mixture model for $P>10$~days, we find $\theta = \{0.042^{+0.018}_{-0.023}, 0.27^{+0.04}_{-0.05}, 0.82^{+0.18}_{-0.09}, 0.10^{+0.05}_{0.09} \}$, entirely consistent with the same model for $P>5$~days, as listed in Table~\ref{tab:distributions}, so that the conclusions about single- and multi-tranet systems are not affected by the exact choice of cut-off period for short-period planets.

\subsubsection{Giant planets} 
\label{sec:giantplanets}

Finally, we investigate planets with $R > 6~R_\oplus$. In our single-tranet sample, there are only five such systems at $P>5$~days, all of them `warm Jupiters', i.e.\ KOI-75b, Kepler-14b, KOI-319b, Kepler-643b, and Kepler-432b. There are three such planets among the multi-tranet systems, i.e.\ Kepler-108b and c, and Kepler-450b. 

For a half-Gaussian distribution, we find $\sigma_\mathrm{single} = 0.31^{+0.11}_{-0.14}$ and $\sigma_\mathrm{multi} = 0.13^{+0.13}_{-0.08}$. If we look at the individual systems, it appears that Kepler-643b and Kepler-432b have a significant and non-zero eccentricity.
For Kepler-108b and c, the orbital eccentricity was measured to be $0.22^{+0.19}_{-0.12}$ and $0.04^{+0.19}_{-0.04}$, respectively, by \cite{vaneylen2015}. An analysis of the TTVs of this system finds an orbital eccentricity of $0.135^{+0.11}_{-0.062}$ and $0.128^{+0.023}_{0.019}$, for planet b and c, respectively, and a significant mutual inclination between the planets \citep{mills2017}. The other systems appear to have orbital eccentricities consistent with zero.
With only a small sample of systems, it is difficult to draw any conclusions. 

Finally, our sample of single-tranet systems consists of three hot Jupiters, i.e.\ $P<5$~days and $R>6~R_\oplus$. For these, we find a half-Gaussian distribution with $\sigma = 0.012^{+0.010}_{-0.011}$, as can be expected from tidal circularization. The multi-tranet sample contains only a single system meeting these criteria (i.e.\ KOI-5b), and its eccentricity posterior is poorly constrained.

\subsubsection{True multiplicity of single-tranet systems}
\label{sec:planetmultiplicity}

Single-tranet systems are not necessarily single-planet systems.  The true multiplicity of systems is unknown, as undetected planets may always reside in the system. However, in some cases TTVs or RVs reveal the presence of additional planets. We have detected clear TTVs for six of the systems in our sample (see Table~\ref{fig:eccsingle_ttvs}), four of which are at $P<5~$days and $R_\mathrm{p}<6~R_\oplus$.

\begin{figure}[!htb]
\centering
\resizebox{\hsize}{!}{\includegraphics{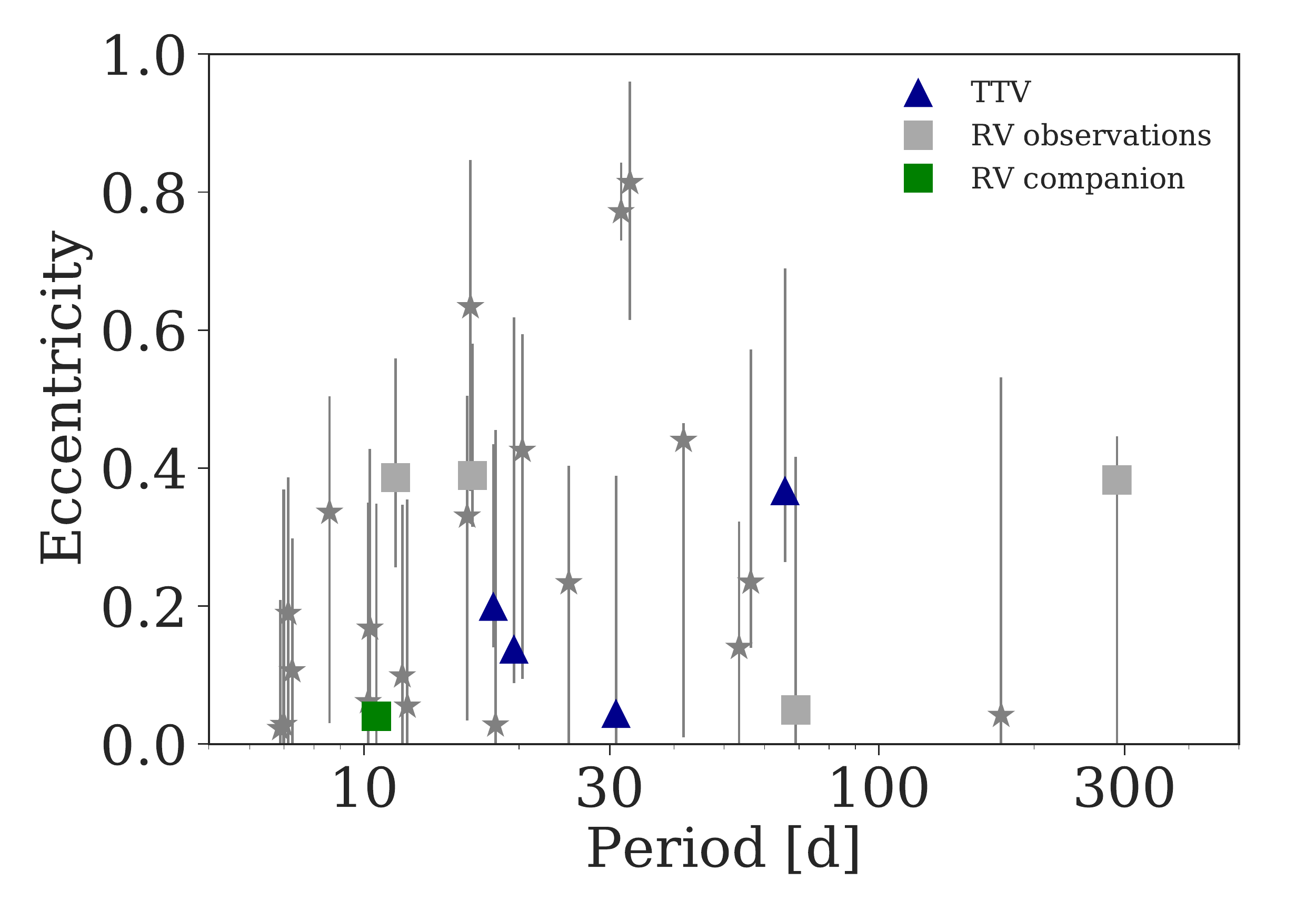}}
\caption{The eccentricity of the single-tranet systems as a function of orbital period, for $P>5$~days and $R_\mathrm{p} < 6~R_\oplus$. Systems with additional detected bodies are flagged in color. The systems with non-transiting companions detected from TTV measurements are shown in blue triangles. Systems with a detected companion through the RV method are shown in green squares. In grey squares, we show the systems where RV follow-up observations have been published, but no companion has been detected, while grey stars show systems where no RV information is available.\label{fig:planet_companions}
}
\end{figure}

The constraints on additional planets from RV monitoring are less clear, because not all systems in our sample have received the same level of RV observations. Three systems at short orbital periods (Kepler-93, Kepler-407, Kepler-408) and three systems with longer orbital periods (Kepler-95, Kepler-96, Kepler-409) have been monitored by \cite{marcy2014}. For two of the short period planets, massive long-period companions were detected, Kepler-93 ($M > 3~M_\mathrm{J}$ and $P > 5~$yr) and Kepler-407 ($M\sin i \approx 5-10~M_\mathrm{J}$ and $P \approx 6-12~$yr), while the other systems show no additional non-transiting planets \citep{marcy2014}. For Kepler-93, \cite{dressing2015} further refine the orbital period and mass of the companion object to be longer than 10 years and more massive than 8.5~$M_\mathrm{J}$, respectively.

TrES-2 has received some RV monitoring with no detected companion \citep{odonovan2006}. HAT-P-7b has a detected long-period companion \citep{winn2009}. HAT-P-11 has a companion \citep{bakos2010,yee2018}. RV observations of Kepler-4 revealed no companion \citep{borucki2010kepler4}.

Kepler-22 received some RV monitoring with a year-long baseline, with no detected signal \citep{borucki2012}. Monitoring of Kepler-7 \citep{latham2010}, Kepler-14 \citep{buchhave2011}, and Kepler-21 \citep{lopezmorales2016} revealed no companions. 

\cite{quinn2015} detect a 406 day period companion to Kepler-432b. Kepler-454 has two non-transiting companions, one with a minimum mass of $4.46 \pm 0.12~M_\mathrm{J}$ in a 524 day orbit, and a second companion with a mass larger than 12.1 $M_\mathrm{J}$ and period longer than 10 years \citep{gettel2016}.

We summarize these observations in Figure~\ref{fig:planet_companions}. There is no obvious pattern linking the presence of TTVs or RV companions to the eccentricity distribution. For the RV observations, at periods longer than 5 days, companions are detected in two systems, one with a significant and one with a low eccentricity, whereas RV monitoring of other systems with both low and higher eccentricities has shown no companions. With the current data, we can therefore not find any correlation linking the detection of a long-period companion to the observed eccentricity, as has been seen for more massive planets \citep{bryan2016}.

In contrast to RVs, TTVs are more sensitive to lower-mass planets at shorter orbital periods, particularly when they are in resonance. To our knowledge, for the systems in our sample only the TTVs of KOI-319b have actually been modeled. \cite{nesvorny2014} find a bimodal solution, with the outer planet having an orbital period of either 80 or 109 days. In both cases, KOI-319b and KOI-319c have low eccentricities, consistent with our findings. In the first case, the mutual inclination between the planets would be $2.37^{+0.91}_{-0.57}$ deg, while in the other it would be $7.3^{+2.3}_{-2.7}$~deg. The low mutual inclination, especially for the first mode, suggests this is a `typical' dynamically cool system, with low eccentricities and low mutual inclinations, where only the inner planet is observed to transit due to its geometry.

The low eccentricities of KOI-75b and Kepler-805 suggest their TTVs could be caused by a low eccentricity, low mutual inclination companion. By contrast, KOI-92b, Kepler-410b, and Kepler-510b show distinctly non-zero eccentricities. An analysis of these TTVs to constrain the mutual inclinations between these planets and their companions would be interesting, but is beyond the scope of this study. If transit duration variations (TDVs) are detected, then these can also be used to constrain mutual inclinations. 

Finally, we note that we detected TTVs in 6/50 single-tranet systems, while \cite{vaneylen2015} detected TTVs in 20/73 multi-tranet systems. Its unclear if this lack of TTVs implies the single-tranet systems have a lower planet multiplicity, or if a detection bias (e.g.\ due to a different orbital separation) is responsible for the lower number of TTV detections.

\section{Comparison with models}
\label{sec:interpretation}

We now investigate the physical processes that cause the observed eccentricity distributions. In Section~\ref{sec:selfexcitation}, we compare our observations to simulations where eccentricities are self-excited; in Section~\ref{sec:outercompanions}, we compare the observations to simulations investigating the role of outer perturbing companions; and finally in Section~\ref{sec:stellarenvironment}, we look into the role of the stellar environment.

\subsection{Self-excitation}
\label{sec:selfexcitation}

During in situ formation of super-Earths, proto-planets can interact gravitationally, a process known as self-stirring.
This process can produce a difference in observed eccentricity distribution between single- and multi-tranet systems, because formation conditions that excite eccentricities also produce wider spacings and larger mutual inclinations, which result in low tranet multiplicity \citep[e.g.][MacDonald et al. in prep.]{moriarty2016,dawson2016}.

A key disk property affecting the observed eccentricity and multiplicity is the solid surface density. A higher solid surface density leads to more proto-planet mergers while the gas disk is still present, causing planets to end up on orbits with smaller eccentricities, tight spacings, and low mutual inclinations. Such dynamically cold systems tend to be observed as multi-tranet systems with low eccentricities.

Another formation parameter that affects the final system architecture is the radial distribution of disk solids. Disks with shallower solid surface density profiles tend to produce systems with fewer transiting planets and higher eccentricities \citep{moriarty2016}. In disks with shallower solid surface density profiles, the embryos at larger semi-major axes are more massive than embryos close to the star. These more massive proto-planets gravitationally stir those that are closer in, producing wider spacings and larger eccentricities and mutual inclinations.

The self-stirring of planets formed in situ leads to eccentricities limited to the ratio of the escape velocity from the surface of the planet to the Keplerian velocity \citep[e.g.][]{goldreich2004,petrovich2014,schlichting2014}, because subsequent close encounters lead to mergers rather than scattering. The final collision tends to reduce the eccentricity further \citep{matsumoto2015}. As a result, typical maximum eccentricities are of order 0.3.

In Figure~\ref{fig:theorymodels}, we compare the eccentricities from an ensemble of in situ formation simulations to those observed. This ensemble consists of 240 simulations. These models are similar to \cite{dawson2016}, but use a distribution of solid surface density normalizations that  are weighted to best match the observed period ratios, the ratios of transit durations between adjacent planets, and multiplicities of the observed \textit{Kepler} transiting planets (MacDonald et al.\ in prep.). In addition, a more accurate planet detection probability is used \citep[KeplerPORTs,][]{burke2017} to transform simulated planetary systems to `observed' transiting planets, rather than the simpler mass and period cut used by \cite{dawson2016}. The solid surface density radial slope is set to $-1.5$. The gas depletion, $d$, relative to the minimum mass solar nebula before the onset of rapid gas disk dispersal, is set to $d=10^4$. For the observations, we plot the best-fit mixture distribution. This distribution is the most straightforward to interpret relative to simulations, as it contains a high-eccentricity and a low-eccentricity component. The latter would be expected for single-tranet systems, because some near-circular multi-planet systems with low mutual inclinations will be \textit{observed} as single-tranet systems, due to the transit geometry. Other observed distributions could be compared to simulations as well, but our goal here is to make a qualitative comparison between observations and simulations, not a quantitative one, so that we plot only one best-fit distribution for simplicity.

The simulated eccentricity distribution depends on the adopted disk parameters. The disk parameters here were optimized to best fit \textit{Kepler} period ratios, ratios of transit durations between adjacent transiting planets, and multiplicities, and do not directly use any observed eccentricity distribution (the ratio of transit duration depends primarily on relative orbital inclinations). As such, other model choices may potentially result in a better match to the observed eccentricity distribution.
As can be seen in Figure~\ref{fig:theorymodels}, the simulations can broadly reproduce the observed eccentricity for single- and multi-tranet systems. Changing the disk parameters would not affect the maximum eccentricities produced by self-stirring or eliminate the trend that the eccentricities in multi-transiting systems are smaller. Changing the disk parameters would alter the shape of the eccentricity distribution below that maximum value for both single and multi-tranet systems, but the observed shape is not well constrained. Therefore, we can interpret the observed eccentricity distribution as being broadly consistent with self-stirring arising from in situ formation but not as validating our choice of disk parameters. High eccentricities, i.e.\ above $\approx 0.3$, cannot be reproduced in this way.

\subsection{Perturbations due to outer companions}
\label{sec:outercompanions}

\begin{figure*}[!htb]
\centering
\resizebox{1\hsize}{!}{\includegraphics{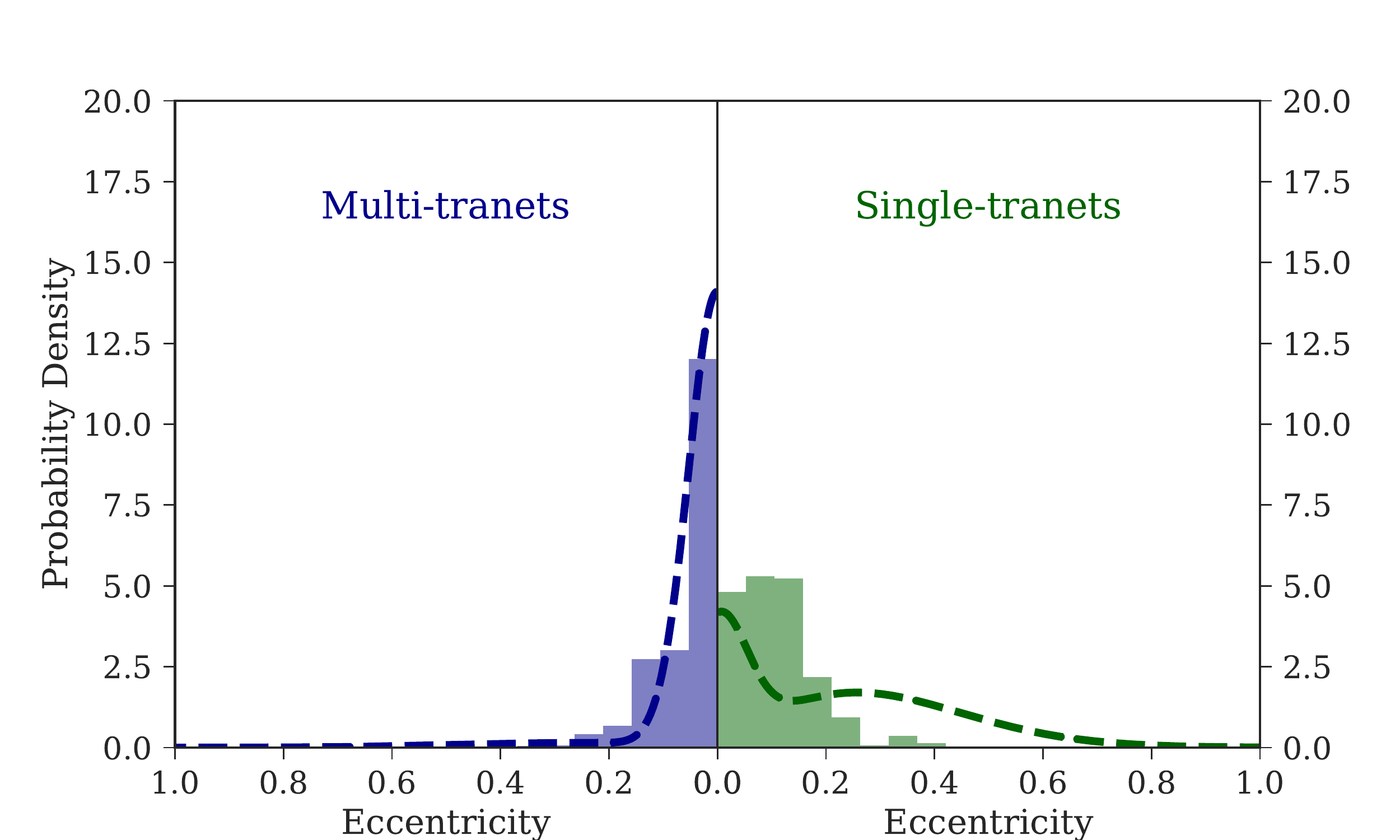}}
\resizebox{1\hsize}{!}{\includegraphics{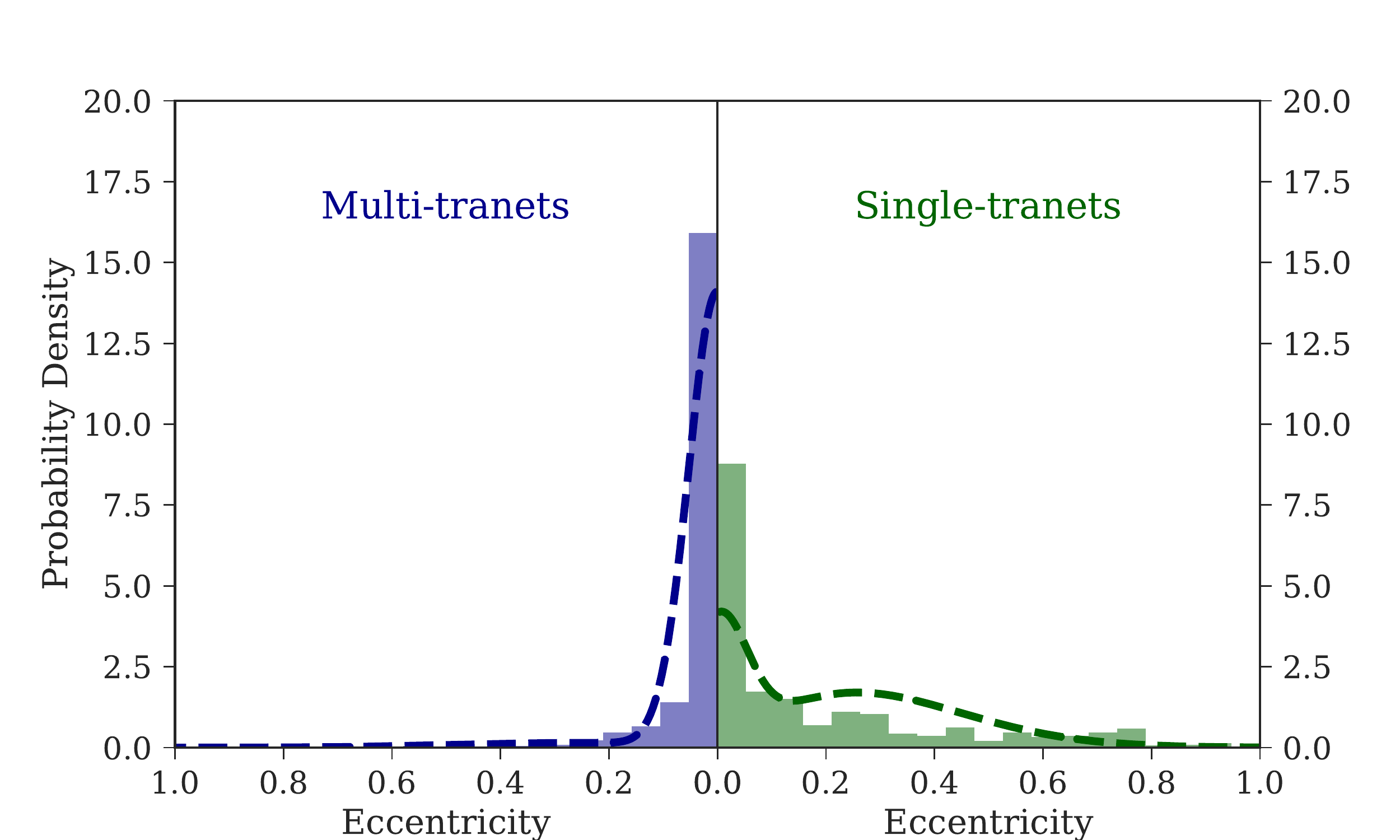}}
\caption{\textbf{Top:} a histogram of eccentricities for single-tranet (green) and multi-tranet systems (blue), from an ensemble of in situ formation simulations in the presence of residual disk gas (see Section~\ref{sec:selfexcitation}). Geometric and detection biases are applied to the simulated sample. The best-fitted mixture distribution to the observation (see Section~\ref{sec:distributionfit}) is overplotted with a dashed line. \textbf{Bottom:} similar to the top plot, but using simulations with outer companion perturbations instead (see Section~\ref{sec:outercompanions}). Such pertubations have a higher-eccentricity tail. The best-fitted mixture distribution to the observation (see Section~\ref{sec:distributionfit}) is again overplotted.}\label{fig:theorymodels}
\vspace{1cm}
\end{figure*}

Several authors have investigated the influence of external companions on compact multi-planet systems containing super-Earths and sub-Neptunes. They found that such perturbations may explain why systems that appear to have fewer transiting planets have a higher eccentricity, because perturbers excite eccentricity as well as increase the mutual inclination between planets, and because compact multi-planet systems may be more resilient to outer perturbations \citep[e.g.][]{lai2017,pu2018}. Similarly, \cite{huang2017} investigated models in which the instability between multiple giant planets at a large distance induces eccentricities of giant planets, and excites eccentricities of close-in super-Earths. The increase of eccentricities in super-Earths are partly due to close encounters with high eccentricity giant planets \citep[see also][]{mustill2017}, and secular interactions with modest eccentric giant planets \citep[see also][]{hansen2017}. If the end result of giant planets interaction gives an eccentricity distribution similar to those seen in radial velocity surveys, the median eccentricity of all survived systems is about 0.2. For those with only one super-Earth remaining in the system, the median eccentricity can be as high as 0.44, with the eccentricity distribution of the single super-Earth almost flat. In Figure~\ref{fig:theorymodels}, we show the expected distribution of eccentricities for observed single and multi-tranet systems following \cite{huang2017}, and compare them with our best-fitted mixture distribution.

It is also possible to excite the eccentricity of inner super-Earths using the inward migration of a single giant planet \citep[see e.g.][Figure 4 therein]{haghighipour2013}. In this scenario, planet embryos can be caught in the mean motion resonance of a migrating giant planet, and undergo dynamic instability. In the specific example demonstrated in \cite{haghighipour2013}, the eccentricity of the super-Earth can reach 0.3 at the end of the simulation. However, it is unclear how often this is the case and how strongly the result depends on the dissipation of the gas disk.

\begin{figure}[!htb]
\centering
\resizebox{\hsize}{!}{\includegraphics{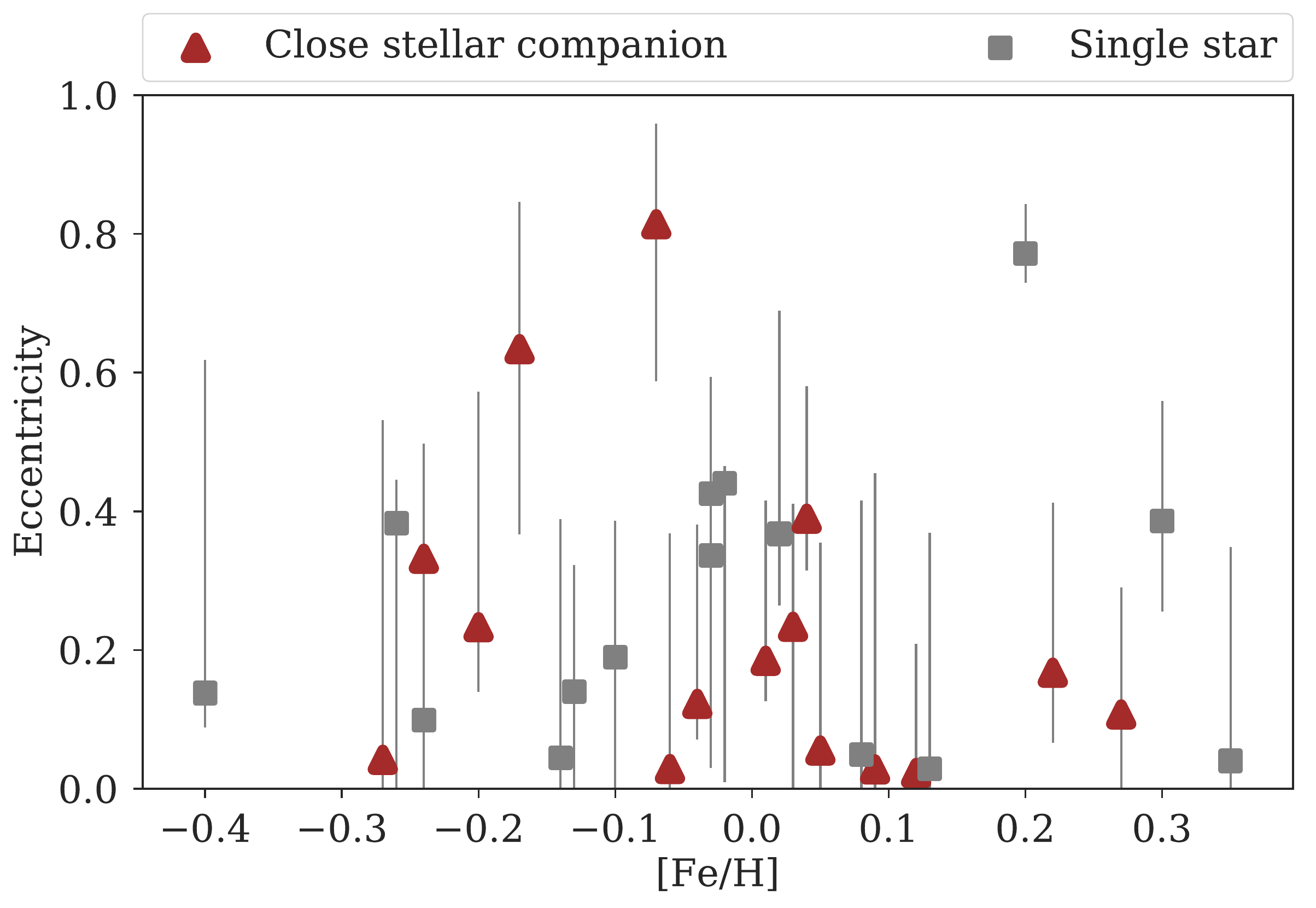}}
\resizebox{\hsize}{!}{\includegraphics{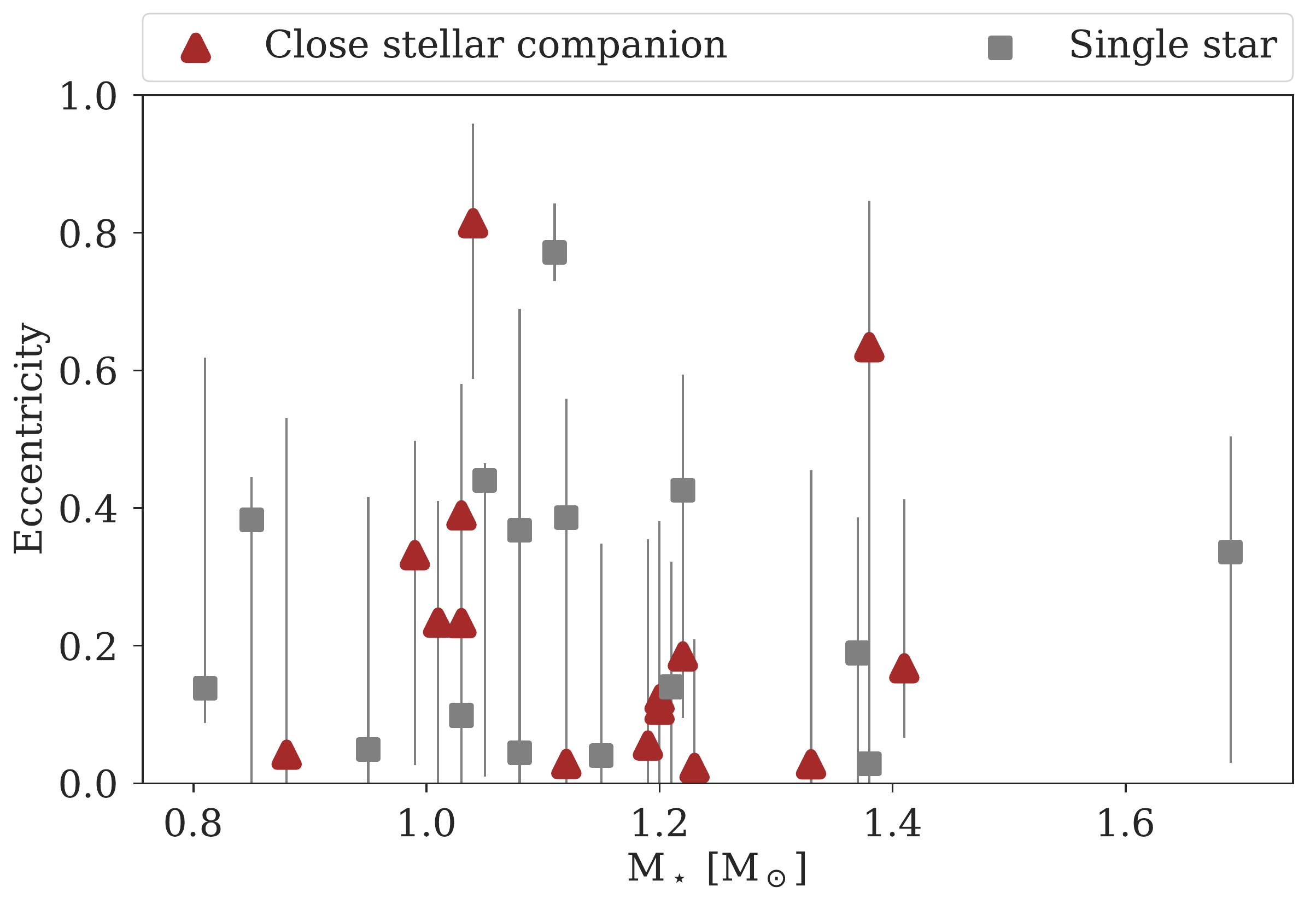}}
\caption{The eccentricity of the single-tranet systems ($R_\mathrm{p} < 6~R_\oplus$ and $P < 5$~days) as a function of stellar metallicity (top) and stellar mass (bottom). The systems are flagged as multi-stellar when they have a detected stellar companion within 4 arcsec, as observed by \cite{furlan2017}. No clear correlation between stellar metallicity and orbital eccentricity, stellar mass and orbital eccentricity, or between multiplicity and orbital eccentricity can be observed in our sample.\label{fig:stellar_companions}
}
\end{figure}

If inner planets are perturbed due to outer giant companions, we can look for evidence of these companions, which may not be transiting. We investigated the true multiplicity of the single-tranet systems in Section~\ref{sec:planetmultiplicity}, but within our limited sample we find no evidence that additional bodies in the system are related to the orbital eccentricity distribution. Since the occurrence of giant planets is correlated with stellar metallicity \citep[e.g.][]{fischer2005}, we can also look for a correlation between orbital eccentricity and stellar metallicity. This is shown in Figure~\ref{fig:stellar_companions}, but we find no evidence of a correlation (i.e. a Spearman rank coefficient of -0.14 with a p-value of 0.47) Similarly, we also show stellar mass and eccentricity in Figure~\ref{fig:stellar_companions}, and find no evidence for a correlation between these parameters (Spearman rank coefficient of -0.13 and p-value of 0.51). More generally, there is evidence that cold Jupiters are common around systems with inner super-Earths \citep{zhu2018}.

\subsection{Perturbations due to stellar environment}
\label{sec:stellarenvironment}

The birth environments of planetary systems may also affect the observed multiplicity of exoplanetary systems. Nascent planetary systems evolve in star cluster environments, where the number densities are a few orders of magnitudes higher than in the Galactic field. Consequently, stellar encounters are frequent, which in turn leads to the excitation of orbital eccentricities and inclination, and even to planet ejections \citep{cai2017a}. 
Through extensive direct $N$-body simulations of multi-planet systems in star clusters, \cite{cai2017b} showed that the multiplicity of a planetary system exhibits an anti-correlation with the mean eccentricity and mean inclination of its planets, and that this anti-correlation is independent of the density of stars in the cluster. 
In this scenario, systems with multiple planets are typically formed in the outskirts of parental clusters where external perturbations are weak and infrequent. In contrast, single-transit systems are dynamically hotter as they were formed in the high-density regions of the cluster, and the strong external perturbations not only lead to the excitation of orbital eccentricities and inclinations, but also reduce the (intrinsic) multiplicity. 

However, while these simulations show the importance of encounters at large orbital periods (i.e.\ several AU), future simulations may clarify whether or not this mechanism can influence the eccentricity and multiplicity of close-in super-Earth systems, like the systems we consider here. It is also possible that encounters have an indirect effect, i.e.\ through their influence on outer planets, which in turn can affect the inner planetary systems, as seen in Section~\ref{sec:outercompanions}.

Similarly, stellar binarity could potentially influence orbital eccentricities. Stellar multiplicity influences the formation and evolution of the protoplanetary disk \citep[e.g.][]{haghighipour2007,andrews2010} and long-period giant planets in binary star systems appear to have a higher eccentricity \citep[e.g.][]{kratter2012,kaib2013}. \cite{mann2017} reported a possible correlation between observed planet multiplicity, eccentricity and stellar multiplicity, in a sample of eight M dwarf systems. 

We check if stellar multiplicity has an influence on the observed eccentricity, using the catalogue compiled by \cite{furlan2017} to check which of the stars in our sample have an observed companion within 4 arcsec of the target star. We mark such systems in Figure~\ref{fig:stellar_companions}. Systems with nearby stellar companions span the whole range of eccentricities, from nearly circular planets to the most eccentric cases, and no obvious difference can be seen between presumed single-star systems and systems with detected stellar companions. A two-sided Kolmogorov-Smirnov statistical test finds a test statistic of $0.27$ and a $p$-value of 0.59, which indicates we cannot rule out the null hypothesis that the eccentricity distribution of planets with and without a stellar companion is the same. Similarly, an Anderson-Darling test for multiple samples results in a test statistic of $-0.51$ and a $p$-value of 0.60, and we cannot reject the hypothesis that the eccentricity distribution of planets with and without a stellar companion is the same. Not all nearby stellar companions are bound to the primary star. \cite{hirsch2017} find that at 1\arcsec, 60-80\% of companions are bound, and that this number decreases for larger separations. However, the chance alignment of a nearby star should not influence the orbital eccentricity, so that chance alignments should not directly affect our statistical test, but may make it more difficult to observe an effect due to stellar multiplicity.
    
In our sample, we find that roughly half (18/35) of the observed planets with orbital periods longer than 5 days have a detected stellar companion. As a robustness check, we investigate the multi-planet sample by \cite{vaneylen2015}, and find similar values: 12/24 have stellar companions detected by \cite{furlan2017}. Roughly 50\% of stars with a nearby stellar companion is a significantly larger fraction than the 30\% which is seen in the full sample of \cite{furlan2017}, but the authors suggest that the true companion fraction may be higher due to sensitivity issues. Previous work suggests that roughly half of Sun-like stars indeed have stellar companions \citep{raghavan2010}. Because our sample consists of relatively bright stars, amenable to asteroseismology, we speculate that this contributes to the detectability of stellar companions, where they may sometimes be missed in other \textit{Kepler} systems.

\section{Discussion}
\label{sec:discussion}

\subsection{Comparison with previous work}

Previous work has noted that \textit{Kepler} single-tranet systems are in some ways dynamically different than multi-tranet systems. \cite{lissauer2011} first noted that a single population that matches the higher multiplicity systems underpredicts the number of tranet systems. By running a range of simulations, tuning the distribution of intrinsic planet multiplicities and mutual inclinations to the observed multi-tranet distribution, \cite{lissauer2011} underpredict the number of observed single-tranet systems, and suggested for the first time that the single-tranet systems contain a population of systems with either a lower intrinsic multiplicity, a higher mutual inclination between planets, or both. They found that a range of possible distributions of mutual inclinations can match the observed multiplicity of double-tranet and triple-tranet systems, this requires mutual inclinations typically lower than 10$^\circ$, but such models underpredict single-tranet systems, and indicate that up to two-thirds of single-tranet systems come from a different population. 

However, these models typically assume only a few total planets (e.g.\ 3-4). \cite{tremaine2012} investigated models that allowed for a very large number of planets (i.e.\ dozens), and find that in such models, a wider range of mutual inclinations can in fact reproduce the observations, up to extreme cases such as an isotropic inclination distribution. 

\cite{hansen2013} studied planet formation by simulating the assembly of planetary embryos for a disk of fixed mass ($20~M_\oplus$ interior to 1~AU) by purely gravitational interactions. They found that these simulations match the characteristics of \textit{Kepler} planets (i.e. inclination, multiplicity, and planet spacing), but underpredict the number of single-planet candidates by about 50\%. \cite{hansen2013} attribute this to unquantified selection effects, an independent process that produces low-multiplicity systems, or additional perturbations which reduce the multiplicity. 

Similarly, \cite{ballard2016} investigated the distribution of single- and multi-tranet systems orbiting M dwarf stars, by running a range of simulations featuring 1-8 planets and a scatter in the mutual inclination of up to 10$^\circ$. They find that $0.53 \pm 0.11$ of the single-tranet systems are either truly single systems, or have additional planets with mutual inclinations larger than those seen in compact multi-planet systems. Along the same lines, \cite{moriarty2016} simulated in situ-planet formation with varying disk solid surface density slopes and normalizations, compared these with observables like the multiplicity and period distribution, and found that high-multiplicity systems make up $24\pm7\%$ of planetary systems orbiting GK-type stars, a lower number than for M-type stars.

While this ``\textit{Kepler} dichotomy'' is often identified in terms of parameters like planet multiplicity, radius, period and period ratio, here we identify a clear difference in orbital eccentricity between single and multi-transiting systems. This provides further evidence of a difference between single-tranet systems and multi-planet systems. However, whether or not this should be interpreted as a `dichotomy' rather than a continuous underlying distribution is unclear, i.e.\ a single mechanism producing a range of orbital eccentricities and mutual inclinations may produce an observed dichotomy between single- and multi-tranets.

\cite{xie2016} modeled the eccentricity distribution of \textit{Kepler} planets using transit durations for a larger but less constrained sample. They found nearly circular multi-planet systems, while single-tranet systems are modeled with a Rayleigh distribution with $\sigma = 0.32$. This difference is observed in this work as well, although we find a lower eccentricity distribution for single-tranet systems when a Rayleigh distribution is used (i.e.\ $\sigma = 0.24^{+0.04}_{-0.04}$ (see Table~\ref{tab:distributions}), which is consistent at the $2\sigma$ level.  

For multi-planet systems which exhibit TTVs, \cite{hadden2014} find a rms eccentricity of $0.018^{+0.005}_{-0.004}$, which can be compared to $\sigma = 0.061^{+0.010}_{-0.012}$ for a Rayleigh distribution fitting the multi-tranet systems -- this value is higher, but includes both systems with and without detected TTVs, suggesting that TTV systems may have a lower typical eccentricity.

At the short orbital period range, our measurements can be compared with eccentricity measurements by \cite{shabram2016}. They determined the eccentricity of short-period (giant) planets by timing the secondary eclipse relative to the primary transit, and found that 90\% of their sample can be characterized with a very small eccentricity ($\approx 0.01$), while the remaining planets come from a sample with a larger dispersion ($0.22$). We investigated eccentricities of short-period planets in Section~\ref{sec:shortperiodplanets}, and find similarly low eccentricities.

Finally, eccentricities have been determined using RVs, but primarily for massive planets. For most systems in our sample, RV measurements are not available. Even when RV observations of small planets lead to mass measurements, the orbital eccentricity can typically not be determined \citep[see e.g.][]{marcy2014}.
\cite{wright2009} observed that systems with masses higher than $1~M_\textrm{J}$ have eccentricities distributed broadly between 0-0.6, while the eccentricities of lower-mass planets are limited to below 0.2.
\cite{mayor2011} see RV eccentricities limited to 0.45 below $30~M_\oplus$, but caution for the trustworthiness of low-eccentricity values in this region of parameter space. The sample investigated here, at $R<6~R_\oplus$, is a region of parameter space for which eccentricity observations are at the edge of what is possible with current RV capabilities.

\subsection{Distinguishing the mechanisms}

We have compared the eccentricities derived here to simulations using several dynamical evolution scenarios (see Section~\ref{sec:interpretation}). In each of these scenarios, higher eccentricities are expected for single tranets because processes that excite eccentricities also excite mutual inclinations and/or widen spacings. How can we then distinguish between them?

Perturbations due to the stellar birth environment or stellar multiplicity likely influence outer giant planets, but there is currently no evidence they influence the close-in small planets investigated here. In the future, simulations focusing on closer-in planetary systems would help investigate whether the birth environment could influence this type of planetary systems. Despite the availability of ground-based high resolution follow-up of the systems investigated here, we find no evidence that the presence of a close stellar companion influences orbital eccentricity (see Figure~\ref{fig:stellar_companions}). 

Eccentricity excitation due to self-gravity of multiple small planets, and eccentricity excitation due to giant outer companions are both able to qualitatively explain the observed eccentricity distribution of single- and multi-tranets. Simulations of each of these effects are both able to broadly match the observed distributions (see Figure~\ref{fig:theorymodels}). Nevertheless, these mechanisms make different predictions at high orbital eccentricities: self-stirring has difficulties to lead to eccentricities above roughly 0.3, while perturbations due to outer companions can lead to planets on highly elliptical orbits. Given our sample size and measurement uncertainties, it is hard to unambiguously determine whether such planets on highly elliptical orbits are present in our sample, although our distribution models (Figure~\ref{fig:theorymodels}) suggest that they do. 

Outer companion perturbations would also imply that single-tranet systems with significant eccentricities are accompanied by giant planets on orbits of approximately $1~$AU \citep{huang2017}. Indirectly, this may lead to a correlation between orbital eccentricity and stellar metallicity, but we find no evidence of this. We also investigated the true multiplicity of single-tranet systems. Although the intrinsic multiplicity is often higher than the tranet multiplicity, as revealed by either RV follow-up or TTV detections, the complete architecture of the systems in our sample remains poorly understood.

If self-excitation is important, the observed multiplicity and eccentricity may depend on stellar type, but our sample is poorly suited to test such predictions: due to the requirement of detecting stellar oscillations, the mass range of stars in our sample is mostly limited to $0.8-1.5~M_\odot$. Other predictions of self-excitation, such as 
a higher bulk density for planets on elliptical orbits at a given mass \citep[for orbital periods beyond the reach of photo-evaporation,][]{dawson2016}, are currently hard to test due to the lack of RV observations for most planets in our sample.

\section{Conclusions}
\label{sec:conclusion}

We conducted a careful modeling of planet transits for systems showing a single transiting planet (single-tranets), and compared those with transit durations from asteroseismology to determine the orbital eccentricity. We compared the eccentricity distribution of single-tranet systems with eccentricities of multi-tranet systems, modeled the observed eccentricity distribution, and compared the distributions to simulations with various planet formation and evolution conditions.

\begin{itemize}
 \item Systems with a single transiting planet exhibit higher average eccentricities than systems with multiple transiting planets. We try different eccentricity distributions, which are summarized in Table~\ref{tab:distributions}. A Rayleigh and half-Gaussian distribution are intuitively simple, while a Beta distribution may be more suitable to use as a prior for future transit modeling work. We also use a mixture model, which points to a significant component ($0.76^{+0.21}_{-0.12}$) with a higher eccentricity for single-tranet systems, while such a component is absent for multi-tranets. 
 \item Regardless of the adopted distributions, there is a clear difference between single-tranet and multi-tranet systems. This difference remains present, even after correcting for the possibility of false positives and the distribution of planet size and orbital period. 
 \item Simulations of an ensemble of systems investigating self-excitation, and simulations investigating the influence of long-period giant companions, can both qualitatively explain our findings. The latter can lead to planets with high eccentricities, while the former can only explain eccentricities up to $\approx 0.3$.
 \item Although several single-tranets show evidence of a higher intrinsic multiplicity, through e.g.\ RV observations or TTV detections, we find no evidence that is related to the orbital eccentricity. We also investigate the role of giant planets in an indirect way, through the stellar metallicity, and find no evidence of a correlation with orbital eccentricity.
 \item Half of the systems in our sample have close companion stars. We find no difference in eccentricity distributions between planets orbiting single stars, and planets orbiting a star with a close stellar companion. 
 \item In Table~\ref{tab:eccsingle_paramtable}, we list the stellar and planetary parameters for this `gold sample' of systems, which may be useful for future studies, e.g.\ this sample clearly shows the presence of the radius gap \citep{vaneylen2017}.
\end{itemize}

The eccentricity distributions derived here may be used as prior information for transit fits of future planet detections, such as those by the upcoming \textit{TESS} mission \citep{ricker2014}. In turn, \textit{TESS} will detect transiting planets orbiting stars brighter than the \textit{Kepler} systems considered in our sample, which may enable a more complete view of the intrinsic architecture of single- and multi-tranet systems. This is likely to help distinguish between the formation and evolution models that can explain the observed orbital eccentricities.

\acknowledgements

We are grateful to the anonymous referee for helpful comments and suggestions which have improved this manuscript. Vincent Van Eylen and Simon Albrecht acknowledge support from the Danish Council for Independent Research, through a DFF Sapere Aude Starting grant No. 4181-00487B. 
Rebekah I. Dawson gratefully acknowledges support from NASA XRP 80NSSC18K0355. Mia S. Lundkvist is supported by The Independent Research Fund Denmark’s Sapere Aude program (Grant agreement no.: DFF–5051-00130). Victor Silva Aguirre acknowledges support from the Villum Foundation (Research grant 10118).  Joshua N. Winn thanks the Heising-Simons Foundation for supporting his work.
This material is based upon work supported by the National Science Foundation Graduate Research Fellowship Program under Grant No. DGE1255832. Any opinions, findings, and conclusions or recommendations expressed in this material are those of the author and do not necessarily reflect the views of the National Science Foundation.
This research has made use of the NASA Exoplanet Archive, which is operated by the California Institute of Technology, under contract with the National Aeronautics and Space Administration under the Exoplanet Exploration Program.
This research made use of the Grendel HPC-cluster for computations. Funding for the Stellar Astrophysics Centre is provided by The Danish National Research Foundation (Grant agreement no.: DNRF106). The research was supported by the ASTERISK project (ASTERoseismic Investigations with SONG and Kepler) and the EXOPLANETBIO project, funded by the European Research Council (Grant agreement no.: 267864, and no. 694513).
  
\bibliographystyle{bibstyle}
\bibliography{references_eccentricities,references_eccentricities_new}

\appendix

\section{Individual planetary systems}
\label{sec:appendix_individual}

\subsection{Short-period planets\label{sec:eccentricity_individual}}

Our sample consists of 15 planets with orbital periods shorter than five days. Three of these systems are hot Jupiters: TrES-2b, HAT-P-7b, and Kepler-7b. They all have eccentricities consistent with circular orbits. The other twelve short-period planets are small. As expected, the majority of these systems has orbits consistent with circularity. There are three exceptions: HAT-P-11b, Kepler-21b, and Kepler-408b. We briefly discuss these systems here.

\subsubsection{HAT-P-11b}

HAT-P-11b is a Neptune-sized planet orbiting its star each 4.9 days. Its eccentricity is found to have a modal value at $0.09$, with a 68\% confidence interval of $[0.06,0.27]$. This is small, but distinctly non-zero, as even at 95\% confidence the eccentricity is found to be within $[0.06,0.59]$, indicating that tides have not had a chance to fully circularize the orbit, or that eccentricity is being pumped into the system. HAT-P-11 is also interesting because the obliquity of the system has been measured and the orbit is found to be oblique \citep{winn2010hatp11,sanchisojeda2011}.

RV observations have independently measured the eccentricity of this planet to be $0.198 \pm 0.046$ \citep{bakos2010}. This value, which is fully consistent with our finding, provides further evidence that the transit duration method can indeed be used to reliably determine eccentricities, even when they are moderate. An eccentric ($e = 0.60 \pm 0.03$) outer planet companion on a long period ($P = 9.3~$years) is present in the system \citep{yee2018}.

\subsubsection{Kepler-21b}

Kepler-21b \citep{howell2012} is a super-Earth orbiting its star each 2.8 days. We find a modal eccentricity value is 0.26, with a 68\% confidence interval spanning [0.11, 0.49]. This is surprising, 
given the short orbital period of the planet. Nevertheless, some caution is warranted: Kepler-21 is a spotted star, complicating the measurement. In addition, the 95\% confidence interval includes a circular orbit, and measures the eccentricity between [0, 0.80]. 

The RV signal of this planet was recently measured by \cite{lopezmorales2016}. The authors determine a mass of $5.08 \pm 1.72 M_\oplus$ and an eccentricity of $0.02 \pm 0.1$, which is consistent with our finding at the $2\sigma$ level. It is therefore unclear if Kepler-21b has a circular orbit, or a mildly eccentric one. A faint stellar companion was discovered near Kepler-21 \citep{ginski2016}.

\subsubsection{Kepler-408b}

Kepler-408b is a small ($0.689 \pm 0.017 R_\oplus$) planet which orbits the star in 2.47 days. 

We find a modal eccentricity value of 0.67, but due to the small size of the planet the error bars are large and at 95\% the orbital eccentricity is consistent with zero. \cite{marcy2014} attempted to constrain the mass of this planet, but could only derive an upper limit of $5 M_\oplus$, and they were therefore unable to constrain the eccentricity.

\subsection{Longer period planets}

Our sample includes 34 systems with a single transiting planet orbiting with a period longer than five days. They display a wider range of eccentricities. In Section~\ref{sec:distributionfit}, we discuss the eccentricity distribution of these systems. Here, we highlight a few individual systems which show a non-zero eccentricity and which may be interesting for follow-up observations.

\subsubsection{Kepler-410A b}

The Kepler-410 system \citep{vaneylen2014} consists of a transiting planet on a 17.8 day orbit, and an additional planet, which is revealed by transit timing variations (see Figure~\ref{fig:eccsingle_ttvs}), as well as a nearby companion star. We find the orbit to be eccentric, with a modal value of 0.2 and a 68\% confidence interval of $[0.14,0.43]$. These values are consistent with the transit duration analysis by \cite{vaneylen2014}, although there a zig-zag-shaped model was used to remove the TTV signal, whereas here we opt for the simpler sinusoidal model.

Asteroseismology has further constrained the inclination of the star to be $i = 82.5 (-2.5, +7.5)^\circ$, making it one of the only single-tranet (but multi-planet) systems for which the obliquity is well-constrained.

\subsubsection{Kepler-95b}

Kepler-95b is a super-Earth orbiting with a period of 11.5 days. With a 68\% confidence interval for eccentricity of $[0.26,0.56]$, the planet's orbit is distinctly eccentric. The planet was confirmed by RV follow-up observations, which measured its mass to be $13.0 \pm 2.9$ M$_\oplus$ \citep{marcy2014}, implying a low planetary density ($1.7 \pm 0.4~$g~cm$^{-3}$) with a large fraction of volatiles. No companion stars were detected. Due to the low RV amplitude, this planet was fitted assuming a circular orbit.

\subsubsection{Kepler-96b (KOI-261b)}

Kepler-96b is a super-Earth with an orbital period of 16.3 days, and we find a relatively large eccentricity with a 68\% confidence interval of $[0.31,0.58]$. Its mass was measured by \cite{marcy2014} to be $8.46 \pm 0.22$ M$_\oplus$. A companion star was also detected by \cite{marcy2014}, but this object is 7 magnitudes fainter and consequentially unlikely to influence our measurement. Due to the small mass of the system, the RV signal was modeled assuming a circular orbit \citep{marcy2014}.

\cite{hirano2012} find the star to be oriented pole-on, by comparing the rotation period from \textit{Kepler} photometry to spectroscopic $v \sin i$ measurements. This suggests that at least in this system, the high obliquity and high eccentricity could have a common origin.

\subsubsection{Kepler-432b}

Kepler-432b is a Jupiter-sized planet orbiting its host star with a period of 52.5 days. The star itself is evolved and has a radius of $4.5$ R$_\odot$. We find an eccentricity of 0.41, with a 68\% confidence interval at $[0.29,0.62]$. This system was the object of three recent and simultaneous studies, which monitored the RV signal of the star. This resulted in planetary mass estimates of $4.87 \pm 0.48$ M$_\textrm{J}$ \citep{ciceri2015}, $5.84 \pm 0.05$ M$_\textrm{J}$ \citep{ortiz2015}, and $5.41^{+0.32}_{-0.18}$ M$_\textrm{J}$ \citep{quinn2015}. These authors also constrained the eccentricity to respectively $0.535 \pm 0.030$, $0.478 \pm 0.004$, and $0.5134^{+0.0098}_{-0.0089}$, measurements which are all fully consistent with our eccentricity measurement. \cite{quinn2015} also measured the stellar parameters using asteroseismology, and used the stellar density to constrain the eccentricity from the transit photometry, in a method that is similar to what is employed here. From this, they find the eccentricity to be $0.507^{+0.039}_{-0.114}$, which is consistent with our measurement. 

Furthermore, it appears that the stellar spin is well aligned with the orbit of the planet \citep{quinn2015}. A long-period planet companion was also detected from RV observations, orbiting with a period of 406 days \citep{quinn2015}.

\subsubsection{KOI-367b}

KOI-367b is a Neptune-sized planet candidate orbiting its host star each 31.6 days. We find a very high eccentricity, with a modal value of 0.77 and a 68\% confidence interval of $[0.73,0.84]$. Adding to the interest of the system, \cite{hirano2012} find that the star's spin-orbit is likely misaligned. Nevertheless, a word of caution is required since this object is an unconfirmed planet candidate. RV follow-up observations of this system may be able to confirm the planet's presence, as well as measure its mass and refine its eccentricity measurement.

\subsubsection{Kepler-643b}

Kepler-643b \citep{morton2016} is a Jupiter-sized planet orbiting its host star in 16.3 days. We find a modal eccentricity value of 0.27, and a 68\% confidence interval of $[0.21,0.49]$, indicating that the orbit of this planetary candidate deviates significantly from circularity.

\newpage

\renewcommand{\arraystretch}{1.6}
\begin{longtable}{llllll|llllllllllll}
 & &	$e$ (mode)	&$e$ (68\%)	&$R_\textrm{p}$ [$R_\oplus$]	&Period [d]		& $M_\star$ [$M_\odot$]		& $R_\star$ [$R_\odot$]		&$\rho_\star$ [g/cm3]&\\
\hline \hline \\[0.2em]
TrES-2&		KOI-1.01&      	$0.01$ &	$[0.0, 0.1]$	&$13.21 \pm 0.28$	&$ 2.47061340 (2) $    	             &$0.97\pm{+0.08}$   &$0.96\pm{+0.02}$	&	$1.548\pm{0.042}$	\\[0.2em]
HAT-P-7&	KOI-2.01&      	$0.01$ &	$[0.0, 0.13]$	&$16.88 \pm 0.26$	&$ 2.20473543 (3) $    	             &$1.55\pm{+0.10}$   &$1.99\pm{+0.03}$	&	$0.279\pm{0.014}$	\\[0.2em]
HAT-P-11&	KOI-3.01&     	$0.09$ &	$[0.06, 0.27]$	&$4.887 \pm 0.065$	&$ 4.88780240 (15)$  	             &$0.86\pm{+0.06}$   &$0.76\pm{+0.01}$	&	$2.743\pm{0.082}$	\\[0.2em]
Kepler-4&	KOI-7.01&      	$0.02$ &	$[0.0, 0.21]$	&$4.22 \pm 0.12$   	&$ 3.21367134 (91)$    	             &$1.09\pm{+0.07}$	&$1.55\pm{+0.04}$	&	$0.410\pm{0.018}$	\\[0.2em]
Kepler-410&	KOI-42.01&  	$0.18$ &	$[0.13, 0.42]$	&$2.786 \pm 0.045$	&$17.833613 (47) $    	             &$1.22\pm{+0.07}$	&$1.35\pm{+0.02}$	&	$0.700\pm{0.030}$	\\[0.2em]
Kepler-93&	KOI-69.01&   	$0.02$ &	$[0.0, 0.18]$	&$1.477 \pm 0.033$	&$ 4.72673930 (86)$    	             &$0.89\pm{+0.07}$	&$0.91\pm{+0.02}$	&	$1.642\pm{0.049}$	\\[0.2em]
        &	KOI-75.01&    	$0.02$ &	$[0.0, 0.18]$	&$10.72 \pm 0.29$    &$105.88162 (75)$     	             &$1.32\pm{+0.07}$	&$2.58\pm{+0.07}$	&	$0.1085\pm{0.0099}$	\\[0.2em]
Kepler-22&	KOI-87.01&   	$0.38$ &	$[0.0, 0.45]$	&$1.806 \pm 0.029$   &$289.8655 (19)$      	             &$0.85\pm{+0.05}$	&$0.83\pm{+0.01}$	&	$2.127\pm{0.059}$	\\[0.2em]
        &	KOI-92.01&    	$0.37$ &	$[0.26, 0.69]$	&$3.00 \pm 0.13$	    &$65.70453 (17)$      	             &$1.08\pm{+0.11}$	&$1.05\pm{+0.03}$	&	$1.316\pm{0.039}$\\[0.2em]
Kepler-7&	KOI-97.01&    	$0.01$ &	$[0.0, 0.15]$	&$17.68 \pm 0.36$	&$ 4.8854862 (12)$   	             &$1.28\pm{+0.07}$	&$1.97\pm{+0.04}$	&	$0.237\pm{0.015}$	\\[0.2em]
Kepler-14&	KOI-98.01&   	$0.04$ &	$[0.0, 0.2]$	&$12.87 \pm 0.26$	&$ 6.7901237 (20)$  	             &$1.34\pm{+0.08}$	&$2.02\pm{+0.04}$	&	$0.228\pm{0.014}$	\\[0.2em]
Kepler-464&	KOI-107.01& 	$0.11$ &	$[0.0, 0.29]$	&$3.44 \pm 0.10$ 	&$ 7.257038 (45)$      	             &$1.2\pm{+0.08} $	&$1.6\pm{+0.04} $	&   $0.411\pm{0.018}$	\\[0.2em]
Kepler-467&	KOI-118.01& 	$0.23$ &	$[0.0, 0.41]$	&$2.26 \pm 0.09$ 	&$24.99337 (21)$       	             &$1.01\pm{+0.07}$	&$1.36\pm{+0.04}$	&	$0.562\pm{0.024}$	\\[0.2em]
Kepler-95&	KOI-122.01&  	$0.39$ &	$[0.26, 0.56]$	&$3.290 \pm 0.094$	&$11.5230844 (97)$     	             &$1.12\pm{+0.08}$	&$1.45\pm{+0.04}$	&	$0.523\pm{0.018}$	\\[0.2em]
Kepler-506&	KOI-257.01& 	$0.02$ &	$[0.0, 0.21]$	&$3.088 \pm 0.082$	&$ 6.8834081 (26)$    	             &$1.23\pm{+0.1}$    &$1.2\pm{+0.03} $	&   $1.006\pm{0.038}$	\\[0.2em]
Kepler-96&	KOI-261.01&  	$0.39$ &	$[0.31, 0.58]$	&$2.647 \pm 0.088$	&$16.2384819 (93)$     	             &$1.03\pm{+0.1}$  	&$0.94\pm{+0.03}$	&	$1.740\pm{0.051}$	\\[0.2em]
        &	KOI-268.01&   	$0.12$ &	$[0.07, 0.38]$	&$3.043 \pm 0.076$   &$110.381$ (n/a)      	             &$1.2\pm{+0.07} $	&$1.36\pm{+0.03}$	&	$0.678\pm{0.030}$	\\[0.2em]
        &	KOI-269.01&   	$0.03$ &	$[0.0, 0.46]$	&$1.549 \pm 0.047$	&$18.01181 (12)$       	             &$1.33\pm{+0.08}$	&$1.45\pm{+0.02}$	&	$0.623\pm{0.034}$	\\[0.2em]
Kepler-454&	KOI-273.01& 	$0.04$ &	$[0.0, 0.35]$	&$2.38 \pm 0.094$	&$10.573754 (12)$   	             &$1.15\pm{+0.11}$	&$1.1\pm{+0.03} $	&   $1.204\pm{0.032}$	\\[0.2em]
Kepler-509&	KOI-276.01& 	$0.44$ &	$[0.01, 0.47]$	&$2.67 \pm 0.20$ 	&$41.746009 (97) $     	             &$1.05\pm{+0.07}$	&$1.19\pm{+0.02}$	&	$0.886\pm{0.025}$	\\[0.2em]
        &	KOI-280.01&   	$0.1$ 	&	$[0.0, 0.35]$	&$2.190 \pm 0.068$	&$11.872877 (11)$      	             &$1.03\pm{+0.09}$	&$1.04\pm{+0.02}$	&	$1.282\pm{0.039}$	\\[0.2em]
Kepler-510&	KOI-281.01& 	$0.14$ &	$[0.09, 0.62]$	&$2.350 \pm 0.077$	&$19.556464 (62)$     	             &$0.81\pm{+0.11}$	&$1.38\pm{+0.04}$	&	$0.432\pm{0.026}$	\\[0.2em]
        &	KOI-288.01&   	$0.17$ &	$[0.07, 0.41]$	&$3.208 \pm 0.055$	&$10.275375 (31)$      	             &$1.41\pm{+0.08}$	&$2.09\pm{+0.03}$	&	$0.2179\pm{0.0099}$	\\[0.2em]
        &	KOI-319.01&   	$0.02$ &	$[0.0, 0.22]$	&$10.36 \pm 0.25$	&$46.15113 (37)$    	             &$1.29\pm{+0.06}$	&$2.08\pm{+0.04}$	&	$0.201\pm{0.013}$	\\[0.2em]
        &	KOI-367.01&   	$0.77$ &	$[0.73, 0.84]$	&$4.72 \pm 0.15$  	&$31.578671 (12)$      	             &$1.11\pm{+0.09}$	&$1.03\pm{+0.03}$	&	$1.438\pm{0.037}$	\\[0.2em]
Kepler-540&	KOI-374.01& 	$0.04$ &	$[0.0, 0.53]$	&$2.947 \pm 0.064$   &$172.70681 (75)$    	             &$0.88\pm{+0.06}$	&$1.15\pm{+0.02}$	&	$0.807\pm{0.031}$	\\[0.2em]
Kepler-643&	KOI-674.01& 	$0.27$ &	$[0.21, 0.49]$	&$11.29 \pm 0.78$	&$16.338888 (59)$  	                 &$1.27\pm{+0.22}$	&$2.78\pm{+0.19}$	&	$0.0831\pm{0.0056}$	\\[0.2em]
        &	KOI-974.01&	    $0.14$ &	$[0.0, 0.32]$	&$2.601 \pm 0.060$	&$53.50593 (20) $      	             &$1.21\pm{+0.08}$	&$1.85\pm{+0.04}$	&	$0.269\pm{0.017}$	\\[0.2em]
Kepler-21&	KOI-975.01&	    $0.26$ &	$[0.11, 0.49]$	&$1.707 \pm 0.043$	&$ 2.7858219 (84)$     	             &$1.27\pm{+0.08}$	&$1.85\pm{+0.03}$	&	$0.285\pm{0.016}$	\\[0.2em]
Kepler-805&	KOI-1282.01&	$0.04$ &	$[0.0, 0.39]$	&$2.611 \pm 0.095$	&$30.8633 (10)$       	             &$1.08\pm{+0.07}$	&$1.59\pm{+0.03}$	&	$0.376\pm{0.017}$	\\[0.2em]
Kepler-432&	KOI-1299.01&	$0.42$ &	$[0.29, 0.63]$	&$14.7 \pm 2.1$  	&$52.5019 (11)$        	             &$1.69\pm{+0.6}$  	&$4.51\pm{+0.63}$	&	$0.0254\pm{0.0042}$	\\[0.2em]
Kepler-815&	KOI-1314.01&	$0.34$ &	$[0.03, 0.5]$	&$4.98 \pm 0.60$ 	&$ 8.57522 (22)$       	             &$1.69\pm{+0.5}$  	&$3.88\pm{+0.43}$	&	$0.0409\pm{0.0042}$	\\[0.2em]
Kepler-407&	KOI-1442.01&	$0.02$ &	$[0.0, 0.3]$	&$1.141 \pm 0.041$	&$ 0.6693127 (20)$	                 &$1.02\pm{+0.07}$	&$1.02\pm{+0.02}$	&	$1.344\pm{0.035}$	\\[0.2em]
Kepler-408&	KOI-1612.01&	$0.67$ &	$[0.47, 0.87]$	&$0.689 \pm 0.017$	&$ 2.465024 (17)$	                 &$1.02\pm{+0.07}$	&$1.21\pm{+0.02}$	&	$0.816\pm{0.025}$	\\[0.2em]
Kepler-907&	KOI-1613.01&	$0.33$ &	$[0.03, 0.5]$	&$1.403 \pm 0.081$	&$15.86631 (51)$	                 &$0.99\pm{+0.08}$	&$1.34\pm{+0.03}$	&	$0.580\pm{0.024}$	\\[0.2em]
Kepler-910&	KOI-1618.01&	$0.03$ &	$[0.0, 0.33]$	&$0.828 \pm 0.049$	&$ 2.364388 (32)$	                 &$1.29\pm{+0.09}$	&$1.5\pm{+0.03} $	&   $0.537\pm{0.024}$	\\[0.2em]
Kepler-911&	KOI-1621.01&	$0.43$ &	$[0.09, 0.59]$	&$2.44 \pm 0.11$ 	&$20.30895 (73)$       	             &$1.22\pm{+0.08}$	&$1.93\pm{+0.04}$	&	$0.239\pm{0.016}$	\\[0.2em]
Kepler-997&	KOI-1883.01&	$0.14$ &	$[0.0, 0.36]$	&$1.304 \pm 0.072$	&$ 2.707295 (36)$                    &$1.09\pm{+0.18}$	&$1.48\pm{+0.07}$	&	$0.468\pm{0.027}$	\\[0.2em]
Kepler-1002&KOI-1890.01&	$0.07$ &	$[0.0, 0.32]$	&$1.609 \pm 0.046$	&$ 4.336422 (31)$     	             &$1.18\pm{+0.07}$	&$1.54\pm{+0.02}$	&	$0.452\pm{0.016}$	\\[0.2em]
Kepler-409&	KOI-1925.01&	$0.05$ &	$[0.0, 0.42]$	&$1.148 \pm 0.048$	&$68.95825 (29)$      	             &$0.95\pm{+0.08}$	&$0.9\pm{+0.02} $	&   $1.824\pm{0.054}$	\\[0.2em]
        &	KOI-1962.01&	$0.81$ &	$[0.59, 0.96]$	&$2.51 \pm 0.14$  	&$32.85861 (58)$       	             &$1.04\pm{+0.07}$	&$1.5\pm{+0.04} $	&   $0.440\pm{0.020}$	\\[0.2em]
        &	KOI-1964.01&	$0.13$ &	$[0.0, 0.37]$	&$0.668 \pm 0.029$	&$ 2.2293226 (85)$     	             &$0.93\pm{+0.11}$	&$0.88\pm{+0.03}$	&	$1.897\pm{0.058}$	\\[0.2em]
Kepler-1219&KOI-2390.01&	$0.63$ &	$[0.37, 0.85]$	&$3.42 \pm 0.37$ 	&$16.1046 (12)$        	             &$1.38\pm{+0.5}$  	&$2.68\pm{+0.25}$	&	$0.101\pm{0.024}$	\\[0.2em]
        &	KOI-2462.01&	$0.06$ &	$[0.0, 0.35]$	&$1.491 \pm 0.083$	&$12.14533 (70)$                     &$1.19\pm{+0.1}$  	&$1.71\pm{+0.04}$	&	$0.332\pm{0.016}$	\\[0.2em]
Kepler-1274&KOI-2545.01&	$0.03$ &	$[0.0, 0.37]$	&$1.441 \pm 0.071$	&$ 6.98156 (21)$	                 &$1.38\pm{+0.07}$	&$2.16\pm{+0.04}$	&	$0.193\pm{0.011}$	\\[0.2em]
Kepler-1298&KOI-2632.01&	$0.19$ &	$[0.0, 0.39]$	&$1.588 \pm 0.089$	&$ 7.12836 (47)$	                 &$1.37\pm{+0.17}$	&$2.16\pm{+0.07}$	&	$0.192\pm{0.024}$	\\[0.2em]
        &	KOI-2706.01&	$0.03$ &	$[0.0, 0.38]$	&$1.797 \pm 0.082$	&$ 3.097597 (22)$	                 &$1.26\pm{+0.18}$	&$1.86\pm{+0.08}$	&	$0.276\pm{0.037}$	\\[0.2em]
Kepler-1392&KOI-2792.01&	$0.04$ &	$[0.0, 0.35]$	&$0.684 \pm 0.052$	&$ 2.128229 (24)$	                 &$0.99\pm{+0.15}$	&$1.3\pm{+0.06} $	&   $0.631\pm{0.030}$	\\[0.2em]
        &	KOI-2801.01&	$0.03$ &	$[0.0, 0.37]$	&$0.870 \pm 0.061$	&$ 6.99180 (16)$	                 &$1.12\pm{+0.17}$	&$1.45\pm{+0.06}$	&	$0.513\pm{0.027}$	\\[0.2em]
Kepler-1394&KOI-2956.01&	$0.04$ &	$[0.0, 0.45]$	&$1.04 \pm 0.11$ 	&$ 3.93800 (32)$	                 &$1.51\pm{+0.21}$	&$1.98\pm{+0.08}$	&	$0.276\pm{0.035}$	\\[0.2em]
        &	KOI-3168.01&	$0.23$ &	$[0.14, 0.57]$	&$0.988 \pm 0.076$	&$56.382 (45)$                       &$1.03\pm{+0.16}$	&$1.55\pm{+0.07}$	&	$0.392\pm{0.025}$	\\[0.2em]
\hline\\[2em]
\caption{Determined parameters for the short-period (top) and long-period planet candidates in our sample. The stellar parameters ($M_\star$, $R_\star$ and $\rho_\star$) are from \cite{lundkvist2016}. \label{tab:eccsingle_paramtable}}
	    \end{longtable}

\end{document}